\documentclass[a4paper,onecolumn]{IEEEtran}
\usepackage{amsfonts}
\usepackage{epsfig}
\usepackage{float}
\usepackage{color}
\usepackage{multicol}
\usepackage{mathcomsteP}
\usepackage{multicol}
\usepackage{subfigure}
\usepackage{cite}
\usepackage{times}

\usepackage{amsmath,amssymb,amsfonts,eucal,latexsym}
\usepackage{graphicx}
\usepackage{pst-plot}
\usepackage{pstricks}
\usepackage{color}

\usepackage{amsthm}

\def \RR{\mathbb R}

\newtheorem{thm}{Theorem}[section]
\newtheorem{cor}[thm]{Corollary}
\newtheorem{lem}[thm]{Lemma}

\theoremstyle{remark}
\newtheorem{rem}[thm]{Remark}
\theoremstyle{definition}

\begin{document}

\title{New inner and outer bounds for the  discrete memoryless cognitive interference channel and some capacity results}

\author{
 \IEEEauthorblockN{Stefano Rini, Daniela Tuninetti, and Natasha Devroye\\}
\medskip
\IEEEauthorblockA {Department of Electrical and Computer Engineering\\
University of Illinois at Chicago\\
Email: \{srini2, danielat, devroye\}@uic.edu}
\thanks{The work of S. Rini and D. Tuninetti was partially funded by NSF under award 0643954.}
}

\maketitle

\begin{abstract}
The cognitive interference channel is an interference channel in which one transmitter is non-causally provided with the message of the other transmitter.
This channel model has been extensively  studied in the past years and capacity results for certain classes of channels have been proved.
In this paper we present new inner and outer bounds for the capacity region of the cognitive
interference channel as well as new capacity results.
Previously proposed outer bounds are expressed in terms of auxiliary random variables for which no cardinality constraint is known.
Consequently it is not possible to evaluate such outer bounds explicitly for a given channel model.
The outer bound we derive is based on an idea originally devised by Sato for the broadcast channel and does not contain auxiliary random variables, allowing it to be more easily evaluated.
The inner bound we derive is the largest known to date and
is explicitly shown to include all previously proposed achievable rate regions.
This comparison highlights which features of the transmission scheme - which includes rate-splitting, superposition coding, a broadcast channel-like binning scheme, and Gel'fand Pinsker coding -  are most effective in approaching capacity.
We next present new capacity results for a class of discrete memoryless channels that we term the ``better cognitive decoding regime" which includes all previously known regimes in which capacity results have been derived as special cases.
Finally, we determine the capacity region of the semi-deterministic cognitive interference channel, in which the signal  at the cognitive receiver is a deterministic function of the channel inputs.

\end{abstract}


\section{Introduction}
\label{sec:intro}

The rapid advancement of wireless technology in the past years has started what some commentators call the ``wireless revolution"
\cite{best2003wireless}.
This revolution envisions a world where one can access telecommunication services on a global scale  without the deployment of local infrastructure.
By increasing the adaptability, communication and cooperation capabilities of wireless devices, it may be possible to realize this revolution.
%
%
Presently, the frequency spectrum is allocated to different entities by dividing it into licensed lots.
Licensed users have exclusive access to their licensed frequency lot or band and cannot interfere with the users in neighboring lots.
%
%
The constant increase of wireless services has led to a situation where new services  have a difficult time obtaining spectrum licenses, and thus cannot be accommodated without discontinuing, or revoking, the licenses of others.
This situation has been termed ``spectrum gridlock" (\cite{goldsmith_survey}) and is viewed as one of the  factors in preventing the emergence of  new services and technologies by entities not already owning significant spectrum licenses.

In recent years, several strategies for overcoming this spectrum gridlock have been proposed \cite{goldsmith_survey}.
In particular,  collaboration among devices and adaptive transmission strategies are envisioned to overcome this spectrum gridlock.
That is, smart and well interconnected devices may cooperate to {\it share} frequency, time and resources to communicate more efficiently and effectively.
The role of information theory in this scenario is to determine ultimate performance limits of a collaborating network.
Given the complexity of this task in its fullest generality, researchers have focussed on simpler models with idealized assumptions.

One of the most well studied and simplest collaborative models is the genie aided cognitive interference channel.
This channel is similar to the classical interference channel:  two senders wish to send information to two receivers. Each transmitter has one intended receiver forming two transmitter-receiver (Tx-Rx) pairs termed the primary and secondary (or cognitive) pairs/users. Over the channel each transmitted message interferes with the other, creating undesired interference at the intended receiver.
This channel model differs from the classical interference channel in the assumptions made about the ability of the transmitters to collaborate:
 collaboration among transmitters is modeled by the idealized assumption that the secondary (cognitive) transmitter has full a-priori (or non-causal) knowledge of the primary message.
This assumption is referred to as genie aided cognition\footnote{This has also been termed ``unidirectional cooperation" or transmission with a ``degraded message sets".}.
The model was firstly posed from an information theoretic perspective in \cite{devroye_IEEE}, where the channel was formally defined and the first achievable rate region was obtained, demonstrating that a cognitive interference channel, employing a form of asymmetric transmitter cooperation, could achieve larger rate regions than the classical interference channel. 
%
The first outer bound  for this channel was derived in  \cite{WuDegradedMessageSet}, together with the first capacity result for a class of channels termed ``very weak interference" in which (in Gaussian noise)  treating interference at the primary user as noise is optimal.
The same achievable rate region was simultaneously derived in \cite{JovicicViswanath06}, where the authors further
characterized the maximum rate achievable by the cognitive user without degrading the rate achievable by the primary user. %
A second capacity result was proved in \cite{Maric_strong_interference} for the so-called  ``very strong interference case", where, without loss of optimality,  both receivers can decode both messages.
%
%
The capacity is also known for the case where the cognitive user decodes both messages \cite{CognitiveInterferenceChannelswithConfidentialMessages} with and without confidentiality constraints.
%

However, the capacity region of the genie aided cognitive radio channel, both for discrete memoryless as well as Gaussian noise channels, remains unknown in general. Tools such as rate-splitting, binning, cooperation and superposition coding have been used to
derive different achievable rate regions.
The authors of \cite{MaricGoldsmithKramerShamai07Eu} proposed an achievable region that encompasses all the previously proposed inner bounds and
derived a new outer bound  using an argument originally devised for the broadcast channel in \cite{NairGamal06}.
A further improvement of the inner bound in \cite{MaricGoldsmithKramerShamai07Eu} is provided in \cite{cao2008} where the authors include a new feature in the transmission scheme allowing the cognitive transmitter to broadcast part of the message of the primary pair.
This broadcast strategy is also encountered in the scheme derived in \cite{maric2008capacityBCCR} for the more general broadcast channel with cognitive relays, which contains the cognitive interference channel as special case. 
%

Many extensions to the cognitive interference channel have been considered. In particular, several papers have addressed the cognitive interference channel's idealized cognition assumption of non-causal,  or a-priori message knowledge at one transmitter.
A more realistic model of cognition is obtained by assuming
a finite (rather than infinite) capacity link(s) between the encoders - termed
the interference channel with conferencing encoders. Under the ``strong interference condition", this channel model reduces to the compound multiple access channel whose capacity was determined  in \cite{MaricYatesKramer07}.
Another non-idealized model for  cognition is the ``causal cognition" model in which the cognitive encoder has access to a channel output and causally learns the primary message.
This models is a special case of the interference channel with generalized feedback of \cite{tuninetti2007interference}, which has considered in \cite{seyedmehdi-achievable} where an achievable scheme using block Markov encoding was derived .
In  \cite{chattarjee-asilomar}, the impact of the knowledge of different codebooks is investigated.

Another natural extension of the cognitive interference channel model is the so called ``broadcast channel with cognitive relays" or ``interference" channel with one cognitive relay".
In this channel model, a cognitive relay in inserted in  a classical interference channel. The cognitive relay has knowledge of the two messages and thus cooperates with the two encoders in the transmission of these two messages.
The model  contains both the interference channel and the cognitive interference channel when removing one of the transmitters and message knowledge (for the interference channel) and thus can reveal
the optimal cooperation trade off between entities in a larger network.
This model was first introduced in \cite{sahin-achievable}, where an achievable rate region was derived.  In \cite{sridharan2008capacity} the authors introduced a larger achievable rate region and derived an outer bound for the sum capacity. In \cite{maric2008capacityBCCR} a yet larger inner bound is derived  by having the cognitive transmitter send a private message to both receivers as in a broadcast channel.

\subsection{Main contributions}

In this paper we establish a series of new results for the discrete memoryless cognitive interference channel.
Section \ref{sec:definitions} introduces the basic definitions and notation.
  Section \ref{sec:Available resuts} summarizes the known results including general inner bounds, outer bounds and capacity in the ``very weak interference" and ``very strong interference" regimes. Our contributions start in Section \ref{sec:new outer bound} and may be summarized as follows:
  \begin{itemize}
\item A new outer bound for the capacity region is presented in Section \ref{sec:new outer bound}: this outer bound is looser than previously derived outer bounds but it does not include auxiliary random variables and thus it can be more easily evaluated.
\item In Section \ref{sec:a new inner bound} we present a new inner bound that encompasses all known achievable rate regions.
\item We show that the newly derived region encompasses all previously presented regions in Section \ref{sec:ComparisonNewAchievableRegion}.
\item We derive the capacity region of the cognitive interference channel in the ``better cognitive decoding" regime in Section \ref{sec:new capacity results}: this regime includes the ``very weak interference" and the ``very strong interference" regimes and is thus the largest set of channels for which capacity is known.
\item Section \ref{sec:semi-det CIFC}  focuses on the semi-deterministic cognitive interference channel in which the output at the cognitive receiver is a deterministic function of the channel inputs. We determine capacity for this channel model by showing the achievability of the outer bound first derived in \cite{WuDegradedMessageSet}.
\item In Section \ref{sec:det CIFC} we consider the deterministic cognitive interference channel: in this case both channel outputs are deterministic functions of the inputs. This channel is a subcase of the semi-deterministic case for which capacity is known. For this channel model we show the achievability of the outer bound proposed in section \ref{sec:new outer bound}, thus showing that this outer bound is tight for this class of channels.
\item The paper concludes with some examples in Section \ref{sec:examples} which provide insight on the role of cognition. We consider two deterministic cognitive interference channel and show the achievability of the outer bound of Section \ref{sec:new outer bound} with transmission strategies over one channel use. The achievable scheme we propose provides interesting insights on the capacity achieving scheme in this channel model -  the extra non-causal message knowledge at one of the transmitters allows a partial joint design of the codebooks and transmission strategies - and is easily appreciated in these simple deterministic models.
\end{itemize}


\section{Channel model, notation and definitions}
\label{sec:definitions}
\begin{figure}[h]
\centering
\includegraphics[width=10 cm]{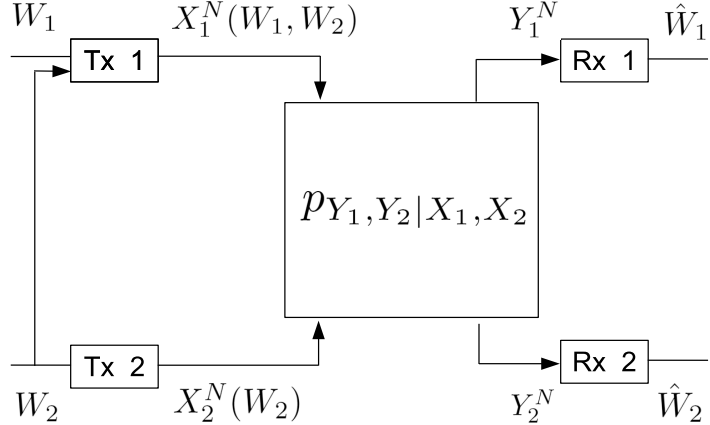}\\
  \caption{The CIFC model. }\label{fig:GeneralModel}
\end{figure}

A two user InterFerence Channel (IFC) is a multi-terminal network with two senders and two receivers. Each transmitter $i$ wishes to communicate a
message $W_i$  to receiver $i\ , \ i=\{1,2\}$. 
In the classical IFC the two transmitters operate independently and have no knowledge of each others' messages. Here we consider
a variation of this set up assuming that transmitter~1 (also called cognitive transmitter), in addition to its own message $W_1$, also knows the message $W_2$ of transmitter 2 (also called primary transmitter).
We refer to transmitter/receiver~1 as the  cognitive pair and to transmitter/receiver~2 as the primary pair.
This model, shown in Figure \ref{fig:GeneralModel}
 is termed the Cognitive InterFerence Channel (CIFC) and is an idealized model for unilateral transmitter cooperation.
The Discrete Memoryless CIFC (DM-CIFC) is a CIFC with finite cardinality input and output alphabets and
a memoryless channel described by the transition probabilities  $p_{Y_1,Y_2|X_1,X_2} (x_1, x_2)$.

Transmitter $i=\{1,2\}$ wishes to communicate a message $W_i$, uniformly distributed on
$[1, \ldots , 2^{N R_i}]$, to receiver $i$ in $N$ channel uses at rate $R_i$.
The two messages are independent.
A rate pair $(R_1,R_2)$ is said to be achievable if there exists
a sequence of encoding functions
\begin{align*}
X_1^N &= X_1^N(W_1, W_2) \\
X_2^N &= X_2^N(W_2),
\end{align*}
and a sequence of decoding  functions
\begin{align*}
\hat{W}_i & = \Wh_i(Y_i^N),  \quad i=\{1,2\}\\
\end{align*}
such  that
\begin{align*}
\lim_{N\goes \infty } \ \ \max_{i=\{1,2 \}}\Pr\lsb  \hat{W}_i \neq W_i  \rsb \to 0.
\end{align*}
The capacity region is defined as the closure of the region of all achievable $(R_1,R_2)$ pairs
 \cite{ThomasCoverBook}.

\section{Existing results for the DM-CIFC}
\label{sec:Available resuts}

We now present  the existing outer bounds and the capacity results available for the DM-CIFC.
%
%
The first outer bound for the  CIFC was obtained in \cite[Thm 3.2]{WuDegradedMessageSet} by the introduction of an auxiliary Random Variable (RV).
\begin{thm}{ \textbf{One auxiliary RV outer bound} of \cite[Thm 3.2]{WuDegradedMessageSet}:} If $(R_1,R_2)$ lies in the capacity region of the DM-CIFC then
\begin{subequations}
\ea{
R_1 &\leq& I(X_1; Y_1|X_2)
\label{eq:outer bound wu R1}\\
R_2 &\leq& I(X_2, U ; Y_2)
\label{eq:outer bound wu R2}\\
R_1+R_2 &\leq& I( X_2,U ;  Y_2)+I(X_1; Y_1| X_2, U),
\label{eq:outer bound wu R1+R2}
}
\label{eq:outer bound wu}
\end{subequations}
taken over the union of distributions that factor as 
$$
p_{U,X_1,X_2}p_{Y_1,Y_2|X_1,X_2}.
$$
\label{thm:outer bound wu}
\end{thm}

Another general outer bound for the capacity region of the CIFC is  provided in  \cite[Thm 4]{MaricGoldsmithKramerShamai07Eu}. This outer bound is derived using an argument originally devised in \cite{NairGamal06} for the Broadcast Channel (BC). The expression of the outer bound is identical to the outer bound in \cite{NairGamal06} but the factorization of the auxiliary RVs differs.

\begin{thm}{ \textbf{BC inspired outer bound} of \cite[Thm. 4 ]{MaricGoldsmithKramerShamai07Eu}:} If $(R_1,R_2)$ lies in the capacity region of the DM-CIFC then
\begin{subequations}
\ea{
R_1 &\leq& I(V, U_1; Y_1)
\label{eq:outer bound maric european R1}\\
R_2 &\leq& I(V, U_2; Y_2)
\label{eq:outer bound maric european R2}\\
R_1+R_2 &\leq& I(V,U_1; Y_1)+I(U_2; Y_2| U_1, V)
\label{eq:outer bound maric european R1+R2 1}\\
R_1+R_2 &\leq& I(V,U_2; Y_2)+I(U_1; Y_1| U_2, V),
\label{eq:outer bound maric european R1+R2 2}
}
\label{eq:outer bound maric european}
\end{subequations}
taken over the union of distributions that factor as 
$$
p_{U_1}p_{U_2}p_{V|U_1,U_2}p_{X_2|U_2,V} p_{X_1| U_1,U_2,V}p_{Y_1,Y_2|X_1,X_2}.
$$
\label{thm:outer bound maric european}
\end{thm}

It is not possible to show in general the containment of the outer bound of Theorem \ref{thm:outer bound wu},``one auxiliary RV outer bound", into the region of Theorem \ref{thm:outer bound maric european}, ``BC inspired outer bound".
%
%

The expression of the outer bound of Theorem \ref{thm:outer bound wu},``one auxiliary RV outer bound", can be simplified in two instances called weak and strong interference.

\begin{cor}{ \textbf{Weak interference outer bound} of \cite[Thm 3.4]{WuDegradedMessageSet}:}

When the condition
\ea{
I(U; Y_2|X_2) \leq I(U; Y_1 | X_2) \ \ \ \forall p_{U,X_1,X_2},
\label{eq:weak interference condition}
}
is satisfied, the outer bound of Theorem \ref{thm:outer bound wu} ,``one auxiliary RV outer bound", can be equivalently expressed as
\begin{subequations}
\begin{align}
R_1 & \leq I(Y_1; X_1|U, X_2)
\label{eq:outer bound weak CIFC R1}
\\
R_2 &\leq I(U,X_2;Y_2),
\label{eq:outer bound weak CIFC R2}
\end{align}
\label{eq:outer bound weak CIFC}
\end{subequations}
taken over the union of all distributions $p_{U,X_1,X_2}$.
\label{thm:outer bound weak}
\end{cor}
We refer to the condition in \reff{eq:weak interference condition} as the ``weak interference condition".

\begin{cor}{ \textbf{Strong interference outer bound} of \cite[Thm 5]{Maric_strong_interference}:}

When the condition
\ea{
I(X_1; Y_1|X_2) \leq I(X_1; Y_2 | X_2)  \ \ \ \  \forall p_{X_1,X_2},
\label{eq:strong interference condition}
}
is satisfied, the outer bound of Theorem \ref{thm:outer bound wu} ,``one auxiliary RV outer bound", can be equivalently expressed as
\begin{subequations}
\begin{align}
R_1 & \leq I(Y_1; X_1| X_2)
\label{eq:outer bound strong CIFC R1}
\\
R_1+R_2 &\leq I(Y_2; X_1,X_2)
\label{eq:outer bound strong CIFC R1+R2}
\end{align}
\label{eq:outer bound strong CIFC}
\end{subequations}
taken over the union of all distributions $p_{X_1,X_2}$.
\label{thm:outer bound strong}
\end{cor}
We refer to the condition in \reff{eq:strong interference condition} as the ``strong interference condition".



The  outer bound of Theorem  \ref{thm:outer bound wu} ,``one auxiliary RV outer bound", may be shown to be achievable in a subset of the ``weak interference" \reff{eq:weak interference condition} and of the ``strong interference" \reff{eq:strong interference condition} conditions.
We refer to these subsets as the ``very strong interference" and ``very weak interference" regimes.

\begin{thm}{ \textbf{Very weak interference capacity} of \cite[Thm. 3.4]{WuDegradedMessageSet} and \cite[Thm. 4.1]{JovicicViswanath06}.}

The outer bound of Corollary \ref{thm:outer bound weak}, ``weak interference outer bound", is the capacity region if
\ea{
I(U; Y_2|X_2) \leq& I(U; Y_1 | X_2) &\nonumber \\
I(X_2; Y_2) \leq & I(X_2; Y_1), & \ \ \ \ \forall p_{U,X_1,X_2} \label{eq:very weak interference condition}.
}
\label{thm:Very weak interference capacity}
\end{thm}
We refer to the condition in \reff{eq:very weak interference condition} as ``very weak interference". In this regime capacity is achieved by having encoder~2 transmit as in a point-to-point channel and encoder~1 perform Gelf`and-Pinsker binning against the interference created by transmitter~2. In a similar spirit, capacity may be obtained in ``very strong interference''.

\begin{thm}{ \textbf{Very strong interference capacity} of \cite[Thm. 5]{Maric_strong_interference}.}
The outer bound of Corollary \ref{thm:outer bound strong}, ``strong interference outer bound", is the capacity region if
\ea{
I(X_1; Y_1|X_2) &\leq& I(X_1; Y_2 | X_2)  \nonumber \\
I(Y_2;X_1, X_2) &\leq& I(Y_1;X_1,X_2),  \ \ \  \ \forall p_{X_1,X_2}.
\label{eq:very strong interference condition}
}
\label{thm: Very strong interference capacity}
\end{thm}
We refer to the condition in \reff{eq:very strong interference condition} as ``very strong interference". In this regime, capacity is achieved by having both receivers decode both messages.

The outer bounds presented in Theorem \ref{thm:outer bound wu}, ``one auxiliary RV outer bound" and \ref{thm:outer bound maric european} , ``BC inspired outer bound", cannot be evaluated in general since they include auxiliary RVs whose cardinality has not yet been bounded.
In the following we propose a new outer bound, looser in general that the outer bound of Theorem \ref{thm:outer bound wu} without auxiliary RVs. This bound is looser than the outer bound of Theorem \ref{thm:outer bound wu},``one auxiliary RV outer bound", in the general case, but it is tight in the ``very strong interference" regime.

%
%
%

\section{A new outer bound}
\label{sec:new outer bound}


\begin{thm}
\label{thm: outer bound CIFC}
 If $(R_1,R_2)$ lies in the capacity region of the DM-CIFC then
\begin{subequations}
\begin{align}
R_1 & \leq I(Y_1; X_1|X_2),
\label{eq:outer bound CIFC R1 us}
\\
R_2 &\leq I(X_1,X_2;Y_2),
\label{eq:outer bound CIFC R2 us}
\\
R_1+R_2& \leq I(X_1,X_2;Y_2)+I(Y_1; X_1| Y_2', X_2),
\label{eq:outer bound CIFC R1+R2 us}
\end{align}
\label{eq:outer bound CIFC us}
\end{subequations}
taken over the union of all distributions  $p_{X_1,X_2}$ and $p_{Y_1,Y_2'|X_1,X_2}$, where $Y_2'$ has the same marginal distribution as $Y_2$, i.e.,
$p_{Y_2'|X_1,X_2}=p_{Y_2|X_1,X_2}$. 

\label{thm:outer bound CIFC us}
\end{thm}
The idea behind this outer bound is to exploit the fact that the capacity region only depends on the marginal distributions $P_{Y_1|X_1,X_2}$ and $P_{Y_2|X_1,X_2}$ because the
receivers do not cooperate.

\begin{proof}
\begin{subequations}
By Fano`s inequality we have that $H(W_i|Y_i^N)\leq N\epsilon_N$, for some $\epsilon_N$ such  that $\epsilon_N \goes 0 $ as $N \goes 0$
for $i\in\{1,2\}$.
The rate of user~1 can be bounded as
\begin{align}
N (R_1 - \epsilon_N)
      &\leq I(W_1;Y_1^N) \nonumber
  \\& \leq I(W_1;Y_1^N|W_2)  \nonumber
  \\& =  I(W_1 , X_1^N(W_1,W_2);Y_1^N|W_2, X_2^N(W_2))  \nonumber
  \\& \leq H(Y_1^N|W_2, X_2^N)-H(Y_1^N|W_2, W_1, X_1^N,X_2^N)  \nonumber
  \\& \leq H(Y_1^N|X_2^N)-H(Y_1^N|W_2, W_1, X_1^N ,X_2^N)  \nonumber
  \\& = H(Y_1^N|X_2^N)-H(Y_1^N| X_1^N ,X_2^N)  \nonumber
\nonumber    \\& =  \sum_{i=1}^N H(Y_{1i}|X_2^N, (Y_1)_1^{i-1})-H(Y_{1i}|X_2^N,X_2^N,(Y_1)_1^{i-1})
\nonumber    \\& \leq  \sum_{i=1}^N H(Y_{1i}|X_{2i})-H(Y_1^N|X_{1i},X_{2i})
\nonumber    \\& = N I(Y_{1T};X_{1T}|X_{2T},T)
\nonumber    \\& = N (H(Y_{1T}|X_{2T},T)-H(Y_{1T}|X_{1T},X_{2T},T))
\nonumber     \\& = N (H(Y_{1T}|X_{2T},T)-H(Y_{1T}|X_{1T},X_{2T}))
\nonumber    \\& \leq N (H(Y_{1T}|X_{2T})-H(Y_{1T}|X_{1T},X_{2T}))
\nonumber    \\& \leq I(Y_{1T}; X_{1T}|X_{2T}),
\label{eq:last passage outer bound R1}
\end{align}

where $T$ is  the time sharing RV, informally distributed over the set $\{1...N\}$ and independent on the other RVs.


The rate of user~2 can be bounded as
\begin{align}
N( R_2-\epsilon_N)
 & \leq   I(Y_2^N;W_2) \nonumber \\
 & \leq  I(Y_2^N;W_2,W_1) \nonumber \\
 & = H(Y_2^N)-H(Y_2^N|W_1,W_2, X_2^N(W_2),X_1^N(W_1,W_2)) \nonumber \\
 & = H(Y_2^N)-H(Y_2^N| X_2^N,X_1^N) \nonumber \\
 & = \sum_{i=1}^N  H(Y_{2i}| (Y_2)_1^{i-1})-H(Y_{2i}|X_1^N,X_2^N,(Y_2)_1^{i-1}) \nonumber\\
 & \leq \sum_{i=1}^N  H(Y_{2i})-H(Y_{2i}|X_{1i},X_{2i}) \nonumber \\
 & \leq N I(Y_{2T};X_{1T},X_{2T}|T) \nonumber \\
 & = N (H(Y_{2T}|T)-H(Y_{2T}|X_{1T},X_{2T},T) \nonumber)\\
  & \leq N (H(Y_{2T})-H(Y_{2T}|X_{1T},X_{2T}) \nonumber)\\
    & \leq I(Y_{2T};X_{1T},X_{2T}).
    \label{eq:last passage outer bound R2}
\end{align}

Next let $Y_2'$ be any RV such that  $P_{Y_2'|X_1,X_2}=P_{Y_2|X_1,X_2}$ but with any joint distribution
$P_{Y_1,Y_2'|X_1,X_2}$.
The sum-rate can then be bounded as
\begin{align}
N (R_1+R_2- 2 N \epsilon_N)
    & \leq I(W_1; Y_1)+I(W_2; Y_2) \nonumber
\\&   \leq I(W_1;Y_1^N|W_2) + I(W_2;Y_2^N)  \nonumber
  \\& \leq I(W_1;Y_1^N, Y_2'^N|W_2) + I(W_2;Y_2^N)\nonumber
  \\& = I(W_2; Y_2^N)+I(W_1;Y_2'^N|W_2)+I(W_1; Y_1^N | Y_2'^N , W_2)\nonumber
  \\& = H(Y_2^N)+\Big(-H(Y_2^N|W_2)+H(Y_2'^N|W_2)\Big) \nonumber
  \\&  \quad -H(Y_2'^N| W_1, W_2)  +H(Y_1^N | Y_2'^N , W_2)-H(Y_1^N | Y_2'^N ,W_1, W_2) \nonumber
  \\& =H(Y_2^N)+H(Y_1^N|W_2,X_2^N,Y_2'^N) \nonumber
  \\&  \quad -H(Y_2'^N| W_1, W_2, X_1^N, X_2^N)-H(Y_1^N |Y_2'^N ,W_1,W_2,X_1^N,X_2^N) \nonumber
  \\& =H(Y_2^N)+H(Y_1^N|W_2,X_2^N,Y_2'^N) \nonumber
  \\&  \quad -H(Y_2^N| X_1^N, X_2^N)-H(Y_1^N |Y_2'^N ,X_1^N,X_2^N) \nonumber
  \\& \leq H(Y_2^N)+H(Y_1^N| X_2^N,Y_2'^N) \nonumber
  \\&  \quad -H(Y_2^N| X_1^N, X_2^N)-H(Y_1^N |Y_2'^N ,X_1^N,X_2^N) \nonumber
\nonumber  \\&\leq  I(Y_{2}^N;X_{1}^N, X_{2}^N)+\sum_{i=1}^{N}H(Y_{1i}| X_2^N,Y_2'^N,(Y_1)_1^{i-1})-H(Y_{1i}| X_1^N,X_2^N,Y_2'^N,(Y_1)_1^{i-1})
\nonumber  \\&\leq  I(Y_{2}^N;X_{1}^N, X_{2}^N)+\sum_{i=1}^{N} H(Y_{1i}| X_{2i},Y_{2i}')-H(Y_{1i}| X_{1i},X_{2i},Y_{2i}')
\nonumber  \\&\leq  I(Y_{2}^N;X_{1}^N, X_{2}^N)+\sum_{i=1}^{N} H(Y_{1i}| X_{2i},Y_{2i}')-H(Y_{1i}| X_{1i},X_{2i},Y_{2i}')
\nonumber  \\&= N \lb I(Y_{2T};X_{1T}, X_{2T})+H(Y_{1T}| X_{2T},Y_{2T}',T)-H(Y_{1T}| X_{1T},X_{2T},Y_{2T}') \rb
\nonumber  \\&\leq  N \lb I(Y_{2T};X_{1T}, X_{2T})+I(Y_{1T}; X_{1T}| X_{2T},Y_{2T}')\rb.
    \label{eq:last passage outer bound R1+R2}
\end{align}

\end{subequations}
\end{proof}


\begin{rem}
The outer bound of Theorem \ref{thm:outer bound CIFC us} contains the outer bound of Theorem \ref{thm:outer bound wu},``one auxiliary RV outer bound".
Indeed, for  a fixed distribution $p_{X_1,X_2}$, $\reff{eq:outer bound wu R1}=\reff{eq:outer bound CIFC R1 us}$ and  $\reff{eq:outer bound wu R2} \leq \reff{eq:outer bound CIFC R2 us}$
since
\pp{
\reff{eq:outer bound wu R2} &= I(Y_2; X_2, U) \\
&\stackrel{(a)}{\leq}
  I(Y_2; X_2, U)+I(Y_2; X_1| U, X_2 )\\
&= I(Y_2; X_1,X_2, U)\\
&= I(Y_2; X_1,X_2)=\reff{eq:outer bound CIFC R2 us},\\
}
where the last equality follows from the Markov chain $U-X_1,X_2-Y_1,Y_2$.

Consider  $Y_2'$ such that $p_{Y_2'|U,X_1,X_2}=p_{Y_2|U,X_1,X_2}$, 
which also implies $p_{Y_2'|U,X_2}=p_{Y_2|U,X_2}$ since
\pp{
p_{Y_2'|U,X_2}&= \f 1 {p_{X_1} }\dint_{|\Ycal_2'|} p_{Y_2'|U,X_1,X_2}  p_{U,X_1,X_2} d X_1 \\
&= \f 1 {p_{X_1} }\dint_{|\Ycal_2|} p_{Y_2|U,X_1,X_2}  p_{U,X_1,X_2} d X_1 \\
&=p_{Y_2|U,X_2},
}

then:
\pp{
\reff{eq:outer bound wu R1+R2}
&= I(Y_2;X_2,U) +I(X_1; Y_1| U, X_2) \\
&= H(Y_2)+H(Y_2| X_1,X_2,U)-H(Y_2| U,X_1,X_2)-H(Y_2|U,X_2) +I(X_1; Y_1| U, X_2) \\
&=I(Y_2; X_1,X_2,U)+H(Y_2'| U,X_1,X_2)-H(Y_2'|U,X_2) +I(X_1; Y_1| U, X_2) \\
& \leq I(Y_2; X_1,X_2)-I(Y_2'; X_1| U,X_2)+I(X_1; Y_1| U, X_2) +I(Y_2';Y_1|U,X_1,X_2) \\
& =I(Y_2; X_1,X_2)-I(Y_2'; X_1| U,X_2)+I(Y_2',X_1; Y_1| U, X_2) \\
&=I(Y_2; X_1,X_2)+I(Y_1;X_1| Y_2', U, X_2) \\
&=I(Y_2; X_1,X_2)+H(Y_1| Y_2', U, X_2)-H(Y_1| Y_2', U, X_1,X_2) \\
& \stackrel{(b)}{\leq}  I(Y_2; X_1,X_2)+H(Y_1| Y_2',  X_2)-H(Y_1| Y_2', X_1,X_2) \\
& =I(Y_2; X_1,X_2)+I(Y_1; X_1| Y_2',X_2)=\reff{eq:outer bound CIFC R1+R2 us}. \\
}
Now the RV $U$ does not appear in the outer bound expression \reff{eq:outer bound CIFC R1+R2 us} and thus we can consider simply the RVs with $p_{\Yt_2|X_1,X_2}=p_{Y_2|X_1,X_2}$ which corresponds to the definition of $Y_2'$ in Theorem \ref{eq:outer bound CIFC us}.

Equality of the outer bounds is verified when conditions $(a)$ and $(b)$ hold with equality, that is when
\pp{
I(Y_2;X_1| U, X_2)=0  \\
I(Y_1;X_1| \Yt_2, U, X_2)=I(Y_1; X_1| \Yt_2,X_2) \ \ \ \forall  p_{U},
}
for a given $\Yt_2$. The first conditions implies the Markov Chain (MC)
$$
Y_2-U,X_2-X_1
$$
and the second condition the MC
$$
Y_1,X_1-\Yt_2 X_2 -U
$$
We currently cannot relate these conditions to any specific class of DM-CIFC. 
\end{rem}

\begin{rem}
The outer bound of Theorem \ref{thm:outer bound CIFC us} reduces to the strong interference outer bound in \reff{eq:outer bound strong CIFC}, in fact
$$
I(Y_1;X_1| X_2) \leq I(Y_2; X_1| X_2 ) \lag \forall  p_{X_1,X_2}
$$
implies
$$
I(Y_1;X_1| Y_2',X_2) \leq I(Y_2; X_1| Y_2' ,X_2 ) \lag \forall  p_{X_1,X_2,Y_2'}.
$$
Now let $Y_2'=Y_2$ to obtain that  $I(Y_1;X_1| Y_2,X_2)=0$
yielding $\reff{eq:outer bound CIFC R1+R2 us}=\reff{eq:outer bound CIFC R2 us}$ so that the two outer bounds coincide.
\end{rem}

\section{A new inner bound}
\label{sec:a new inner bound}

As the DM-CIFC encompasses classical interference, multiple-access and broadcast channels, we expect to see a combination of their achievability proving techniques surface in any unified scheme for the CIFC. Our achievability scheme employs the following classical techniques:  

\smallskip

\noindent $\bullet$ {\bf Rate-splitting.} We employ a rae-splitting technique similar to that in Han and Kobayashi's scheme of \cite{Han_Kobayashi81} for the interference-channel, also employed in the DM-CIFC regions of \cite{MaricGoldsmithKramerShamai07Eu, devroye_IEEE, JiangXinAchievableRateRegionCIFC}. While rate-splitting may be useful in general,  is not necessary in the very weak \cite{WuDegradedMessageSet} and very strong \cite{MaricUnidirectionalCooperation06} interference regimes of \reff{eq:very weak interference condition} and \reff{eq:very strong interference condition}. \\
$\bullet$  {\bf Superposition-coding.} Useful in multiple-access and broadcast channels \cite{ThomasCoverBook}, in the DM-CIFC the superposition of  private messages on top of common ones, as in \cite{MaricGoldsmithKramerShamai07Eu, JiangXinAchievableRateRegionCIFC}, is known to be capacity achieving in very strong interference \cite{MaricUnidirectionalCooperation06}. \\
$\bullet$  {\bf Binning.} Gel'fand-Pinsker coding \cite{GelFandPinskerClassic}, often simply referred to as binning, allows a transmitter to ``cancel" (portions of) the interference known to be experienced at a receiver.  Binning is also used by Marton in deriving the largest known achievable rate region \cite{MartonBroadcastChannel} for the discrete memoryless broadcast channel.

We now present a new achievable rate region for the DM-CIFC which
generalizes all the known achievable rate regions presented in
\cite{MaricGoldsmithKramerShamai07Eu,WuDegradedMessageSet, JiangXinAchievableRateRegionCIFC, DevroyeThesis, cao2008}
and \cite{jiang-achievable-BCCR}.

\begin{thm}{\textbf{ The $\Rcal_{RTD}$ region.}}
\label{thm:our achievable region}
A rate pair $(R_1,R_2)$ such that
\ea{
R_1 &=& R_{1c}+R_{1pb}, \nonumber \\
R_2 &=& R_{2c}+R_{2pa}+R_{2pb}. \label{eq:rate split}
}
is achievable for the DM-CIFC if $(R_{1c}',R_{1pb}',R_{2pb}',R_{1c},R_{1pb},R_{2c},R_{2pa},R_{2pb})\in\RR^8_+$ satisfies:
\begin{subequations}
\ea{
R'_{1c}                          &=& I(U_{1c};X_{2}|U_{2c})
\label{eq:our achievable region R0"}
\\
R'_{1c}  +R'_{1pb}             &\geq& I(U_{1pb}; X_{2}| U_{1c}, U_{2c})  + I(U_{1c};X_{2}|U_{2c})
\label{eq:our achievable region R0"+R1"}
\\
R'_{1c}  +R'_{1pb} +R'_{2pb} &\geq&  I(U_{1pb}; X_{2}, U_{2pb}| U_{1c},U_{2c})           +I(U_{1c}; X_{2}|U_{2c})
\label{eq:our achievable region R1"+R2"}
\\
R_{2c}+R_{2pa}+(R_{1c}+R'_{1c}  )+(R_{2pb}+R'_{2pb}) &\leq& I(Y_2; U_{2pb},U_{1c},X_{2},U_{2c}) +I(U_{1c}; X_{2}| U_{2c})
\label{eq:our achievable region Ed2:1}
\\
       R_{2pa}+(R_{1c}+R'_{1c}  )+(R_{2pb}+R'_{2pb}) &\leq& I(Y_2; U_{2pb},U_{1c},X_{2}| U_{2c}) +I(U_{1c}; X_{2}| U_{2c})
\label{eq:our achievable region Ed2:2a}
\\
       R_{2pa}                   +(R_{2pb}+R'_{2pb}) &\leq& I(Y_2; U_{2pb},X_{2}| U_{1c},U_{2c}) +I(U_{1c}; X_{2}| U_{2c})
\label{eq:our achievable region Ed2:2b}
\\
               (R_{1c}+R'_{1c}  )+(R_{2pb}+R'_{2pb}) &\leq& I(Y_2; U_{2pb}, U_{1c}| X_{2},U_{2c})+I(U_{1c}; X_{2}| U_{2c})
\label{eq:our achievable region Ed2:3a}
\\
                                  (R_{2pb}+R'_{2pb}) &\leq& I(Y_2; U_{2pb}| U_{1c},X_{2},U_{2c})
\label{eq:our achievable region Ed2:3b}
\\
R_{2c}+(R_{1c}+R'_{1c}  )+(R_{1pb}+R'_{1pb} ) &\leq& I(Y_1; U_{1pb},U_{1c},U_{2c}),
\label{eq:our achievable region Ed1:1}
\\
       (R_{1c}+R'_{1c}  )+(R_{1pb}+R'_{1pb} ) &\leq& I(Y_1; U_{1pb},U_{1c}|U_{2c}),
\label{eq:our achievable region Ed1:2}
\\
                          (R_{1pb}+R'_{1pb} ) &\leq& I(Y_1; U_{1pb}|U_{1c},U_{2c}),
\label{eq:our achievable region Ed1:3}
}
\label{eq:our achievable region}
\end{subequations}
for some input distribution
\pp{
p_{Y_1,Y_2,X_1,X_2,U_{1c},U_{2c},U_{2pa},U_{1pb},U_{2pb}}
&=p_{U_{1c},U_{2c},U_{2pa},U_{1pb},U_{2pb},X_1,X_2} p_{Y_1,Y_2|X_1,X_2}.
}
\end{thm}

\begin{rem}
Moreover:
\begin{itemize}
  \item \reff{eq:our achievable region Ed2:1} can be dropped when $R_{2c}=R_{2pa}=R_{2pb}=R_{2pb}'=0$;
  \item \reff{eq:our achievable region Ed2:2a} can be dropped when $R_{2pa}=R_{2pb}=R_{2pb}'=0$;
  \item \reff{eq:our achievable region Ed2:3a} can be dropped when $R_{2pb}=R_{2pb}'=0$;
  \item \reff{eq:our achievable region Ed1:1} can be dropped when $R_{1c}=R_{1c}'=R_{1pb}=R_{1pb}'=0$,
\end{itemize}

since they correspond to the event that a common message from the non-intended user is incorrectly decoded.
This event is not  an error event if no other intended message is incorrectly decoded.
\end{rem}

\begin{proof}
The meaning of the RVs in Theorem~\ref{thm:our achievable region} is as follows.
Both transmitters perform superposition of two codewords: a common one (to be decoded at both decoders)
and a private one (to be decoded at the intended decoder only). In particular:
\begin{itemize}
  \item Rate $R_1$ is split into $R_{1c}$ and $R_{1pb}$ and conveyed through the RVs $U_{1c}$ and $U_{1pb}$, respectively.
  \item Rate $R_2$ is split into $R_{2c}$, $R_{2pa}$ and $R_{2pb}$ and conveyed through the RVs $U_{2c}, X_{2}$ and $U_{2pb}$, respectively.
  \item $U_{2c}$  is the common  message of transmitter~2. The subscript ``c" stands for ``common".
  \item $X_{2}$ is the private message of transmitter~2 to be sent by transmitter~2 only.
        It superimposed to $U_{2c}$. The subscript ``p" stands for ``private" and the subscript ``a" stands for ``alone".
  \item $U_{1c}$  is the common message of transmitter~1. It is superimposed to $U_{2c}$ and - conditioned on $U_{2c}$ - is
         binned against $X_{2}$.
  \item $U_{1pb}$ and $U_{2pb}$ are private messages of transmitter~1 and transmitter~2, respectively,
         and are sent by transmitter~1 only. They are binned against one another conditioned on $U_{2c}$,
         as in Marton's achievable rate region for the broadcast channel~\cite{MartonBroadcastChannel}.
         The subscript ``b" stands for ``broadcast".
  \item $X_{1}$ is finally superimposed to all the previous RVs and transmitted over the channel.
\end{itemize}

A graphical representation of the encoding scheme of Theorem \ref{thm:our achievable region} can be found in Figure \ref{fig:encodingScheme}. The formal description of the proposed encoding scheme is as follows:

\begin{figure*}
\epsfig{figure=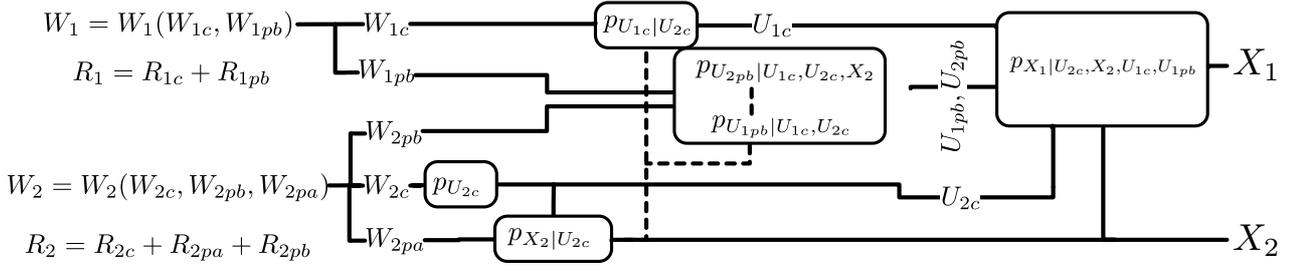, width=17cm}
 \caption{The achievability encoding scheme of Thm. \ref{thm:our achievable region}. The ordering from left to right and the distributions demonstrate the codebook generation process. The dotted lines indicate binning. We see rate splits are used at both users, private messages $W_{1pb}, W_{2pa}, W_{2pb}$ are superimposed on common messages $W_{1c}, W_{2c}$ and $U_{1c}$, is binned against $X_2$ conditioned on $U_{2c}$, while $U_{1pb}$ and $U_{2pb}$ are binned against each and $X_2$ other in a Marton-like fashion (conditioned on other subsets of RVs).  }
 \label{fig:encodingScheme}
\end{figure*}

\subsection{Rate splitting}
Let $W_1$ and $W_2$ be two independent RVs uniformly distributed on $[1...2^{N R_1}]$  and $[1...2^{N R_2}]$  respectively.
Consider splitting the messages as follows:
\pp{
W_1 = (W_{1c},W_{1pb}), \\
W_2 = (W_{2c},W_{2pb},W_{2pa}),
}
where the messages $W_{i}$, $ i\in\{1c,2c,1pb,2pb,2pa\}$, are all independent and
uniformly distributed on $[1...2^{N R_i}]$, so that the rate are
\pp{
R_1=R_{1c}+R_{1pb},\\
R_2=R_{2c}+R_{2pa}+R_{2pb}.
}
\subsection{Codebook generation}
Consider a distribution $p_{U_{1c}, U_{2c}, X_{2},U_{1pb},U_{1pb},X_1,X_2}$. The codebooks are generated as follows:
\begin{itemize}
  \item Select uniformly at random $2^{N R_{2c}}$ length-$N$ sequences
  $U_{2c}^N(w_{2c})$, $w_{2c} \in [1...2^{N R_{2c}}]$,
  from the typical set $T_\epsilon^N(p_{U_{2c}})$.

  \item For every $w_{2c} \in [1...2^{N R_{2c}}]$, select uniformly at random $2^{N R_{2pa}}$ length-$N$ sequences
  $X_{2}^N(w_{2c},w_{2pa})$, $w_{2pa}\in [1... 2^{N R_{2pa}}]$,
  from the typical set $T_\epsilon^N(p_{X_{2},U_{2c}}|U_{2c}^N(w_{2c}))$.

  \item For every $w_{2c} \in [1...2^{N R_{2c}}]$, select uniformly at random $2^{N (R_{1c}+R'_{1c}  )}$ length-$N$ sequences
  $U_{1c}^N(w_{2c},w_{1c},b_0)$, $w_{1c} \in [1...2^{N R_{1c}}]$ and $b_0 \in [1...2^{N R'_{1c}  }]$,
  from the typical set $T_\epsilon^N(p_{U_{1c}U_{2c}}|U_{2c}^N(w_{2c}))$

  \item For every $w_{2c} \in [1...2^{N R_{2c}}]$, $w_{2pa}\in [1... 2^{N R_{2pa}}]$,
  $w_{1c} \in [1... 2^{N R_{1c}}]$ and $b_0 \in [1... 2^{N R'_{1c}  }]$,
  select uniformly at random $2^{N(R_{2pb}+R'_{2pb})}$ length-$N$ sequences
  $U_{2pb}^N(w_{2c},w_{2pa},w_{1c},b_0, w_{2pb},b_2)$, $w_{2pb}\in [1... 2^{N R_{2pb}}]$ and $b_2 \in [1...2^{N R'_{2pb}}]$,
  from the typical set $T_\epsilon^N(p_{U_{2pb},U_{2c},U_{1c},X_{2}}|U_{2c}^N(w_{2c}), X_{2}^N(w_{2c},w_{2pa}), U_{1c}^N(w_{2c},w_{1c},b_0))$.

  \item For every $w_{2c} \in [1...2^{N R_{2c}}]$,
  $w_{1c} \in [1...2^{N R_{1c}}]$ and $b_0 \in [1...2^{N R'_{1c}  }]$,
  select uniformly at random $2^{N(R_{1pb}+R'_{1pb} )}$ length-$N$ sequences
  $U_{1pb}^N(w_{2c},w_{1c},b_0, w_{1pb},b_1)$, $w_{1pb}\in [1... 2^{N R_{1pb}}]$ and $b_1 \in [1...2^{N R'_{1pb} }]$,
  from the typical set
\pp{
   T_\epsilon^N(p_{U_{1pb},U_{2c},U_{1c}}|U_{2c}^N(w_{2c}), U_{1c}^N(w_{2c},w_{1c},b_0)).
   }
  \item For every $w_{2c} \in [1... 2^{N R_{2c}}]$, $w_{2pa}\in [1... 2^{N R_{2pa}}]$,
  $w_{1c} \in [1... 2^{N R_{1c}}]  $, $b_0 \in [1:2^{N R'_{1c}  }]$,
  $w_{1pb}\in [1... 2^{N R_{1pb}}]$, $b_1 \in [1:2^{N R'_{1pb} }]$,
  $w_{2pb}\in [1... 2^{N R_{2pb}}]$, $b_2 \in [1:2^{N R'_{2pb}}]$,
  let the channel input $X_1^N(w_{2pa},w_{2c},w_{1c},b_0, w_{1pb},b_1, w_{2pb},b_2)$ be any length-$N$ sequence
  from the typical set
  \pp{T_\epsilon^N(p_{X_{1},U_{2c},U_{1c},X_{2},U_{2pb},U_{1pb}}
  |U_{2c}^N(w_{2c}), X_{2}^N(w_{2c},w_{2pa}), U_{1c}^N(w_{2c},w_{1c},b_0),U_{2pb}^N(w_{2c},w_{2pa},w_{1c},b_0, w_{2pb},b_2),\\ U_{1pb}^N(w_{2c},w_{1c},b_0, w_{1pb},b_1)).}

  \end{itemize}

\subsection{Encoding}
\label{sec:encoding}
Given the message $w_2=(w_{2c},w_{2pb},w_{2pa})$,
encoder~2 sends the codeword $X_2^N (w_{2c},w_{2pa})$.

Given the message $w_2=(w_{2c},w_{2pb},w_{2pa})$ and the message
$w_1=(w_{1c},w_{1pb})$, encoder~1 looks for a triplet $(b_0,b_1,b_2)$ such that:
\begin{align*}
 &(U_{2c}^N(w_{2c}),
  X_{2}^N(w_{2c},w_{2pa}),
   U_{1c}^N(w_{2c},w_{1c},b_0),
  U_{1pb}^N(w_{2c},w_{1c},b_0,w_{1pb},b_1),
  U_{2pb}^N(w_{2c},w_{1c},b_0,w_{2pb},b_2))
\\&\in T_\epsilon^N(
  p_{U_{2c},X_{2},U_{1c},U_{1pb},U_{2pb}}).
\end{align*}
If no such triplet exists, it sets $(b_0,b_1,b_2)=(1,1,1)$.
If more than one such triplet exists, it picks one uniformly at random
from the found ones. For the selected $(b_0,b_1,b_2)$, encoder~1 sends
$X_1^N(w_{2pa},w_{2c},w_{1c},b_0, w_{1pb},b_1, w_{2pb},b_2)$.

Since the codebooks are generated iid according to
\begin{align}
p^{\rm (codebook)}=p_{U_{2c}}\,p_{X_{2}|U_{2c}}\,p_{U_{1c}|U_{2c}}\, p_{U_{2pb}|U_{2c},U_{1c},X_{2}}\,p_{U_{1pb}|U_{2c},U_{1c}}
\label{eq:pdf codebook}
\end{align}
but the encoding forces the actual transmitted
codewords to look as if they were generated iid according to
\begin{align}
p^{\rm (encoding)}=p_{U_{2c}}\,p_{X_{2}|U_{2c}}\,p_{U_{1c}|U_{2c},X_{2}}\, p_{U_{2pb}|U_{2c},U_{1c},X_{2}}\,p_{U_{1pb}|U_{2c},U_{1c},X_{2},U_{2pb}},
\label{eq:pdf enccoding}
\end{align}
We expect the probability of encoding error to depend on
\[
E\left[\frac{p^{\rm (encoding)}}{p^{\rm (codebook)}}\right]
=E\left[\frac{p_{U_{1c}|U_{2c},X_{2}}\,p_{U_{1pb}|U_{2c},U_{1c},X_{2},U_{2pb}}}{p_{U_{1c}|U_{2c}}\,p_{U_{1pb}|U_{2c},U_{1c}}}\right]
=I(U_{1c}; X_{2}|U_{2c})+I(U_{1pb}; X_{2}, U_{2pb}| U_{2c},U_{1c}).
\]

\subsection{Decoding}
Decoder~2 looks for a unique tuple $(w_{2c},w_{2pa},w_{2pb})$ and some $(w_{1c},b_0,b_2)$ such that
\[
  (U_{2c}^n(w_{2c}),
  X_{2}^n(w_{2c},w_{2pa}),
   U_{1c}^n(w_{2c},w_{1c},b_0),
  U_{2pb}^n(w_{2c},w_{1c},b_0,w_{2pb},b_2),Y_2^n)
  \in T_\epsilon^n(
  p_{U_{2c},X_{2},U_{1c},U_{2pb},Y_2}).
\]

Depending on which messages are wrongly decoded at decoder~2,
the transmitted sequences and the received $Y_2^n$ are generated iid according to
\begin{align}
p_{2|\star}
\triangleq p_{U_{2c}}\,p_{X_{2}|U_{2c}}\,p_{U_{1c}|U_{2c}}\, p_{U_{2pb}|U_{2c},U_{1c},X_{2}}\,
p_{Y_2|\star},
\label{eq:pdf tx rx 2}
\end{align}
where ``$\star$" indicates the messages decoded correctly.
However, the actual transmitted sequences and the received $Y_2^n$
considered at decoder~2 look as if they were generated iid according to
\begin{align}
p_{2}
\triangleq p_{U_{2c}}\,p_{X_{2}|U_{2c}}\,p_{U_{1c}|U_{2c},X_{2}}\, p_{U_{2pb}|U_{2c},U_{1c},X_{2}}\,
p_{Y_2|U_{2c},U_{1c},X_{2},U_{2pb}}.
\label{eq:pdf dec 2}
\end{align}
Hence we expect the probability of error at decoder~2 to depend on terms of the type
\begin{align}
I_{2|\star}
=E\left[\log\frac{p_{2}}{p_{2|\star}}\right]
=E\left[\log\frac{p_{U_{1c}|U_{2c},X_{2}}\,p_{Y_2|U_{2c},U_{1c},X_{2},U_{2pb}}}{p_{U_{1c}|U_{2c}}\,p_{Y_2|\star}}\right]
=I(U_{1c}; X_{2}|U_{2c})+I(Y_2; U_{2c},U_{1c},X_{2}, U_{2pb}| \star).
\label{eq:mutual info 2}
\end{align}


Decoder~1 looks for a unique pair $(w_{1c},w_{1pb})$ and some $(w_{2c},b_0,b_1)$ such that
\[
  (U_{2c}^n(w_{2c}),
   U_{1c}^n(w_{2c},w_{1c},b_0),
  U_{1pb}^n(w_{2c},w_{1c},b_0,w_{1pb},b_1),Y_1^n)
  \in T_\epsilon^n(
  p_{U_{2c},U_{1c},U_{1pb},Y_1}).
\]

Depending on which messages are wrongly decoded at decoder~1,
the transmitted sequences and the received $Y_1^n$ are generated iid according to
\begin{align}
p_{1|\star}\triangleq p_{U_{2c}}\,p_{U_{1c}|U_{2c}}\, p_{U_{1pb}|U_{2c},U_{1c}}\,
p_{Y_1|\star},
\label{eq:pdf tx rx 1}
\end{align}
where ``$\star$" indicates the messages decoded correctly.
However, the actual transmitted sequences and the received $Y_1^n$
considered at decoder~1 look as if they were generated iid according to
\begin{align}
p_{1}\triangleq p_{U_{2c}}\,p_{U_{1c}|U_{2c}}\, p_{U_{1pb}|U_{2c},U_{1c}}\,
p_{Y_1|U_{2c},U_{1c},U_{1pb}}.
\label{eq:pdf dec 1}
\end{align}
Hence we expect the probability of error at decoder~1 to depend on terms of the type
\begin{align}
I_{1|\star}
=E\left[\log\frac{p_{1}}{p_{1|\star}}\right]
=E\left[\log\frac{p_{Y_1|U_{2c},U_{1c},U_{1pb}}}{p_{Y_1|\star}}\right]
=I(Y_1; U_{2c},U_{1c},U_{1pb}| \star).
\label{eq:mutual info 1}
\end{align}


The error analysis  is found in Appendix \ref{app:error analysis RTD}.
\end{proof}

\subsection{Two step binning}
\label{sec: two step binning}
It is also possible to perform binning in a sequential manner. First, $U_{1c}$ is binned against $X_1$, and then $U_{1pb}$ and $U_{2pb}$ are binned against each other conditioned on $(U_{2c},U_{1c})$  and $(U_{2c},X_2,U_{1c})$ respectively.
With respect to the encoding operation of the previous section,  this affects Section \ref{sec:encoding} as follows:

Given the message $w_2=(w_{2c},w_{2pb},w_{2pa})$ and the message
$w_1=(w_{1c},w_{1pb})$, encoder~1 looks for $b_0$
such that
\begin{align*}
 &(U_{2c}^N(w_{2c}),
  X_{2}^N(w_{2c},w_{2pa}),
   U_{1c}^N(w_{2c},w_{1c},b_0), \\&\in T_\epsilon^N(p_{U_{2c},X_{2},U_{1c}}).
\end{align*}

If no such $b_0$ exists, it sets $b_0=1$.
If more than one such $b_0$ exists, it picks one uniformly at random.
For the selected $b_0$, encoder~1 looks for $(b_1,b_2)$  such that:
\begin{align*}
 &(U_{2c}^N(w_{2c}),
  X_{2}^N(w_{2c},w_{2pa}),
   U_{1c}^N(w_{2c},w_{1c},b_0),
  U_{1pb}^N(w_{2c},w_{1c},b_0,w_{1pb},b_1),
  U_{2pb}^N(w_{2c},w_{1c},b_0,w_{2pb},b_2))
\\&\in T_\epsilon^N(
  p_{U_{2c},X_{2},U_{1c},U_{1pb},U_{2pb}}).
\end{align*}
If no such $(b_1,b_2)$ exists, it sets $(b_1,b_2)=(1,1)$.
If more than one such $(b_1,b_2)$ exists, it picks one uniformly at random
from the found ones.

For the selected $(b_0,b_1,b_2)$, encoder~1 sends
$X_1^N(w_{2pa},w_{2c},w_{1c},b_0, w_{1pb},b_1, w_{2pb},b_2)$.

The next lemma states the condition under which  this two step encoding procedure is successful with high probability.

\begin{lem}
\label{thm:two step binnig}

The two-step binning encoding  procedure of Section \ref{sec: two step binning} is successful if
\eas{
R'_{1c}       &\geq& I(U_{1c}; X_2 |U_{2c}),
\label{eq: two step binning R"1c}\\
R'_{1pb}     &\geq& I(U_{1pb}; X_2 |U_{2c},U_{1c}),
\label{eq: two step binning R"1pb}\\
R'_{1pb} +R'_{2pb}&\geq& I(U_{1pb};  X_2, U_{2pb}| U_{2c},U_{1c}).
\label{eq: two step binning R1pb+R"2pb}
}{\label{eq:two step binning}}

\end{lem}

The proof of the lemma is found in Appendix \reff{app:proof two step binnig}.

\begin{rem}
Since the binning rate \reff{eq:our achievable region R0"} of Theorem \ref{thm:our achievable region}  can be taken with equality, the two step binning has the same performance as joint binning. In fact, by setting $\reff{eq: two step binning R"1c}$ to hold with equality, we obtain the equality between the binning rate expression of the joint binning and the two step binning.
\end{rem}

\begin{figure}[h]
\epsfig{figure=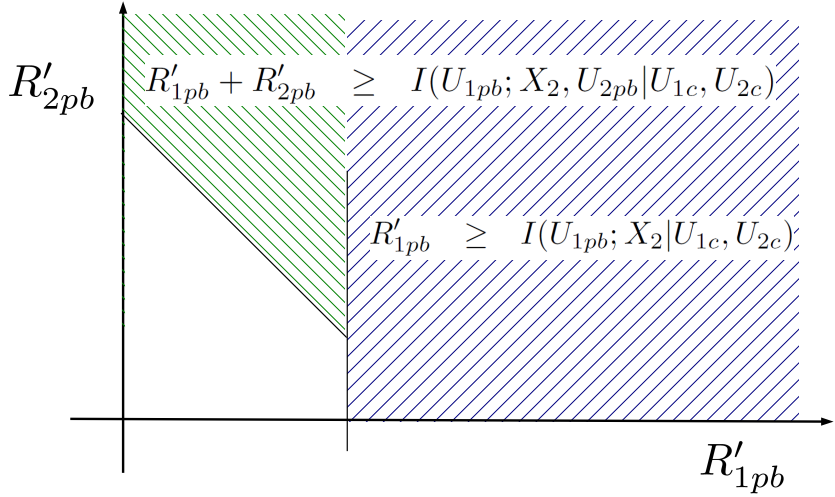, width=12cm}
 \caption{The region of the admissible binning rates $R_{1pb}$ and $R_{2pb}$ in Theorem \ref{thm:our achievable region}. }
 \label{fig:BinningPlot}
\end{figure}

A plot of the permissible  binning rates $R_{1pb}$ and $R_{2pb}$ is depicted in Figure \ref{fig:BinningPlot}.

\section{Comparison with existing achievable rate regions}
\label{sec:ComparisonNewAchievableRegion}

We now show that the  region of Theorem \ref{thm:our achievable region} contains all other known achievable rate
regions for the DM-CIFC.
Showing inclusion of the rate regions \cite[Thm.2]{biao2009Arxiv},  \cite[Thm. 1]{DevroyeThesis} and \cite[Thm. 4.1]{jiang-achievable-BCCR}
is sufficient to demonstrate the largest known DM-CIFC region, since the region of  \cite[Thm.2]{biao2009Arxiv} (first presented in \cite{cao2008})  is shown (in \cite{biao2009Arxiv}) to contain those of
 \cite[Thm. 1]{MaricGoldsmithKramerShamai07Eu} and  \cite{JiangXinAchievableRateRegionCIFC}.

 \subsection{Devroye et al.'s region  \cite[Thm. 1]{DevroyeThesis}}
 \label{sec:devroye}
In Appendix \ref{sec:devroye thesis achievable region} we show that the region of  \cite[Thm. 1]{DevroyeThesis} ${\cal R}_{DMT}$, is contained in our new region  ${\cal R}_{RTD}$ along the lines:

\noindent
$\bullet$ We make a correspondence between the random variables and corresponding rates of ${\cal R}_{DMT}$ and ${\cal R}_{RTD}$.  \\
$\bullet$ We define new regions  $\Rcal_{DMT}\subseteq \Rcal_{DMT}^{out}$ and $\Rcal_{RTD}^{in} \subseteq \Rcal_{RTD}$ which are easier to compare: they have  identical input distribution decompositions and similar rate equations. \\
$\bullet$ For any fixed input distribution, an equation-by-equation comparison leads to  $\Rcal_{DMT} \subseteq \Rcal_{DMT}^{out} \subseteq  \Rcal_{RTD}^{in} \subseteq  \Rcal_{RTD}$.

\subsection{Cao and Chen's region \cite[Thm. 2]{biao2009Arxiv}}
\label{sec:biao}

The region in \cite[Thm. 2]{biao2009Arxiv} uses a similar encoding structure as that of $\Rcal_{RTD}$ with two exceptions:

1) The binning is done sequentially rather than jointly as in $\Rcal_{RTD}$ leading to binning constraints (43)--(45) in
\cite[Thm. 2]{biao2009Arxiv} as opposed to \eqref{eq:our achievable region R0"}--\eqref{eq:our achievable region R1"+R2"}
 in Thm.\ref{thm:our achievable region}. Notable is that both schemes have adopted a Marton-like binning scheme at the
 cognitive transmitter, as first introduced in the context of the CIFC in \cite{cao2008}.

 2) While the cognitive messages are rate-split in identical fashions, the primary message is split into 2 parts in
  \cite[Thm. 2]{biao2009Arxiv} ($R_1=R_{11}+R_{10}$, note the reversal of indices)
   while we explicitly split the primary message into three parts $R_2 = R_{2c}+R_{2pa}+R_{2pb}$.
   In Appendix   \ref{sec:biao}
    we show that the region of \cite[Thm.2]{biao2009Arxiv}, denoted as ${\cal R}_{CC} \subseteq {\cal R}_{RTD}$ in  two steps:

\noindent $\bullet$  We first show that we may WLOG set $U_{11} = \emptyset$ in \cite[Thm.2]{biao2009Arxiv}, creating a new region $R_{CC}'$.

\noindent $\bullet$  We next make a correspondence between our RVs and those of \cite[Thm.2]{biao2009Arxiv} and  obtain identical
 regions. 

\subsection{Jiang et al.'s region \cite[Thm. 4.1]{jiang-achievable-BCCR}}
\label{sec:jiang BCCR}
The scheme originally designed for the more general broadcast channel with cognitive relays (or interference-chanel with a cognitive relay) may be tailored/reduced to derive a region for the cognitive interference channel.  This scheme also incorporates a broadcasting strategy. However, the common messages are created independently instead of having the common message from transmitter~1 superposed to the common message from transmitter~2. The former choice introduces more rate constraints than the latter and allows us to show inclusion in $\Rcal_{RTD}$ after equating random variables.
The proof of the containment of the achievable region of \cite[Thm. 4.1]{jiang-achievable-BCCR} in $\Rcal_{RTD}$  is found in Appendix \ref{sec:Jiang BCCR region}.

\section{New capacity results for the DM-CIFC}
\label{sec:new capacity results}

%
%
We now  look at the expression  of the outer bound  \cite[Thm. 3.1]{WuDegradedMessageSet}  to gain insight into potentially capacity achieving achievable schemes.  In particular we look at the expression of the corner points of the outer bound region for a fixed $p_{U,X_1,X_2}$ and try to interpret the RVs as private and common messages to be decoded at the transmitter side.  We then consider an achievable scheme inspired by these observations and  show that schemes achieve capacity for a particular class of channels. This class of channels contains the ``very strong'' and the ``very weak'' interference regimes and thus corresponds to the largest class of channels for which capacity is currently known.

The outer bound region of \cite[Thm. 3.1]{WuDegradedMessageSet} has at most two corner points where both $R_1$ and $R_2$ are non zero:
\ea{
(R_1^{out \  (a)} ,R_2^{out \ (a)})=\lb I(Y_1; X_1|U,X_2),I(Y_2;U,X_2) \rb
\label{eq: (r1a,r2a)} \\
(R_1^{out \ (b)} ,R_2^{out \ (b)})=\lb I(Y_1; X_1|U,X_2)+I(Y_2;U,X_2)-\De , \De \rb
\label{eq: (r1a,r2a)}\\
    \lag \De=[I(Y_2; U, X_2)- I(Y_1;U| X_2)]^+ ,
    \nonumber
}
since
\pp{
R_2^{out \ (a)} &= \min\{I(Y_2;U,X_2), I(Y_2;U,X_2)+I(Y_1; X_1|U,X_2)\}
          = I(Y_2;U,X_2), \\
R_1^{out \ (a)} &= \min\{I(Y_1; X_1|U,X_2), I(Y_1; X_1|X_2)\}
          = I(Y_1; X_1|U,X_2),
}
and
\pp{
R_2^{out \ (b)} &= \min\{I(Y_2;U,X_2),I(Y_2;U,X_2)+I(Y_1; X_1|U,X_2)-I(Y_1; X_1|X_2)\}\\
          &=[I(Y_2;U,X_2)+\min\{0,I(Y_1; X_1|U,X_2)-I(Y_1; X_1,U|X_2)\}]^+\\
          &=[ I(Y_2;U,X_2)-I(Y_1; U| X_2) ]^+ \triangleq \De, \\

R_1^{out \ (b)} &\leq \min\{I(Y_1; X_1|X_2), I(Y_2;U,X_2)+I(Y_1; X_1|U,X_2)\} \\
          &= I(Y_1; X_1|U,X_2)+I(Y_2;U,X_2)-\max\{I(Y_2;U,X_2)-I(Y_1; U|X_2),0 \}\\
          & =I(Y_1; X_1|U,X_2)+I(Y_2;U,X_2)-\De. \\
}
Proving the achievability of both these corner points for any $p_{U,X_1,X_2}$ shows capacity by a simple time sharing argument.

We can now look at the corner point expression and try to draw some intuition on the achievable schemes that can possibly achieve these rates.
For the corner point $(R_1^{(a)},R_2^{(a)})$ we can interpret $(U,X_2)$ as a  common message from transmitter~2 to receiver~2 that is also decoded at receiver~1. $X_1$ is superposed to $(U,X_2)$ since the decoding of $X_1$ follows the one of $(U,X_2)$  at decoder~2.

The corner point $(R_1^{out \ (b)},R_2^{out \ (b)})$ has two possible expressions:

1) If $I(Y_1; U|X_2) \leq I(Y_2;U,X_2)$ we have that
\ea{
(R_1^{out \ (b)'},R_2^{out \ (b)'})=\lb I(Y_1; X_1,U| X_2), I(Y_2;U,X_2)-I(Y_1; U| X_2)  \rb,
\label{eq: (r1b",r2b")}
}
which  suggests that $X_2$ is again the common primary message and the cognitive message is divided into a  public and private part, $U$ and $X_1$ respectively.

2) If $I(Y_1; U|X_2) >  I(Y_2;U,X_2)$ we have that
\ea{
(R_1^{out \ (b)"},R_2^{out \ (b)"})=\lb I(Y_2;U,X_2)+I(Y_1; X_1,U| X_2), 0  \rb
\label{eq: (r1b",r2b")}.
}
In this case the outer bound has only one corner point  where both rates are non zero. Note that we can always achieve the point
$$
(R_1^{in \ (b)"},R_2^{in \ (b)"})=\lb I(Y_1; X_1, U|X_2), 0  \rb
$$
by having transmitter~2 send a known signal.
In this case we have $R_2^{out \ (b)"}=R_2^{in \ (b)"}$ and  $R_1^{out \ (b)"} \leq R_1^{in \ (b)"}$ since
\pp{
I(Y_1; X_1, U|X_2) \geq I(Y_2;U,X_2)+I(Y_1; X_1,U| X_2)\\
I(Y_1; U|X_2) >  I(Y_2;U,X_2).
}
So in this case showing the achievability of the point in equation \reff{eq: (r1a,r2a)} is sufficient to show capacity.

Guided by these observations, we consider a scheme that has only the components $U_{2c},U_{1c}$  and $U_{1pb}$. That is,
the primary message $\om_2$ is common and the cognitive message $\om_1$ is split into a private and a public message.
With this scheme we are able to extend the capacity results in the ``very weak interference" of Theorem \ref{thm:Very weak interference capacity} and the ``very strong interference" of Theorem \ref{thm: Very strong interference capacity}.
This scheme coincides with  the scheme of \cite{JiangXinGarg07} which  achieves capacity if the cognitive receiver is required  to decode  both messages (with and without the secrecy constraint).
%

\begin{thm}{\textbf{Capacity in the ``better cognitive decoding" regime.}}

When the following condition holds
\ea{
I(Y_1; X_2, U) \geq I(Y_2; X_2,U) \lag \forall p_{X_1,X_2,U},
}
the capacity region of the DM-CIFC is given by region in \eqref{eq:outer bound wu}.
\end{thm}
\begin{proof}
Consider the achievable rate region of Theorem \ref{thm:our achievable region} when setting
\pp{
X_1= U_{1pb}\\
X_2=U_{2c}=U_{2pb}\\
}
so that
\pp{
R_2= R_{2c}\\
R_{2pa}=R_{2pb}=0\\
R_{1c}'=R_{1pb}'=R_{2pb}'=0.
}
In the resulting scheme, the message from transmitter~2 to receiver~2 is all common while the message from transmitter~1 to receiver~1 is split into common and private parts.
The achievable region of this sub-scheme is:
\begin{subequations}
\ea{
R_{2}+R_{1c} &\leq& I(Y_2; U_{1c},X_2)\\
R_{2}+R_{1c}+R_{1pb}&\leq& I(Y_1; U_{1c},X_2)\\
R_{1c}+R_{1pb}&\leq& I(Y_1;U_{1c},X_1|U_{2c})\\
R_{1pb}&\leq& I(Y_1;X_1|X_2,U_{1c}).
}
\end{subequations}
By applying Fourier-Motzkin elimination \cite{lall-advanced} we obtain the achievable rate region
\eas{
R_1 &\leq&  I(Y_1;U_{1c},X_1|X_2)
\label{eq:S6 R1}\\
R_2 &\leq& I(Y_2;U_{1c},X_2)
\label{eq:S6 R2}\\
R_1+R_2 &\leq& I(Y_2;U_{1c},X_2)+I(Y_1;X_1|X_2,U_{1c})
\label{eq:S6 R1+R2-1}\\
R_1+R_2 &\leq& I(Y_1; X_2,U_{1c},X_1).
\label{eq:S6 R1+R2-2}
}
By letting $U_{1c}=U$ we see that \reff{eq:outer bound wu R1} matches \reff{eq:S6 R1},  \reff{eq:outer bound wu R2} matches \reff{eq:S6 R2}, \reff{eq:outer bound wu R1+R2} matches \reff{eq:S6 R1+R2-1}, and \reff{eq:S6 R1+R2-2} is redundant when
$$
I(Y_1; X_2,X_1,U) \geq  I(Y_2; U,X_2)+I(Y_1;X_1|X_2,U),
$$
or equivalently when
\ea{
I(Y_1;U,X_2)      \geq  I(Y_2; U,X_2).
\label{eq:better cognition condition}
}

\end{proof}
We term the condition in equation \reff{eq:better cognition condition} ``better cognitive decoding" since decoder~1 has a higher mutual information between  its received channel output and the RVs $U$ and $X_2$ than the primary receiver.

\begin{rem}
The ``better cognitive decoding" in \reff{eq:better cognition condition} is looser than both the ``very weak interference" condition of
\reff{eq:very weak interference condition}
and the ``very strong interference" condition of
\reff{eq:very strong interference condition}.
In fact  summing the two equations of condition \reff{eq:very weak interference condition} we have
\pp{
I(U; Y_1 | X_2) +I(X_2; Y_1)  \geq I(U; Y_2|X_2) + I(X_2; Y_2)
\IFF I(Y_1; U,X_2) \geq I(Y_2; U,X_2)
}
which corresponds to condition \reff{eq:better cognition condition}.
Similarly by summing the two equation of condition \reff{eq:very strong interference condition} we obtain
\pp{
I(Y_1;X_1, X_2) +I(X_1; Y_2 | X_2) &\geq& I(Y_2;X_1, X_2)+I(X_1; Y_1|X_2)    \IFF \\
I(Y_1;X_1, X_2) -I(X_1; Y_1 | X_2) &\geq& I(Y_2;X_1, X_2)-I(X_1; Y_2|X_2)    \IFF \\
I(Y_1;X_1, X_2,U) -I(X_1; Y_1 | X_2) &\geq& I(Y_2;X_1, X_2,U)-I(X_1; Y_2|X_2)    \IFF \\
I(Y_1;X_2,U)&\geq& I(Y_2;X_2,U)\\
}
which again corresponds to condition \reff{eq:better cognition condition}.

Since both \reff{eq:very weak interference condition} and \reff{eq:very strong interference condition} imply the \reff{eq:better cognition condition},
we conclude that \reff{eq:better cognition condition} is more general than the previous two.

The scheme that achieves capacity in very weak interference is obtained by setting  $U_{1c}=X_{2}$ so that all the cognitive message is private and the primary message is common.
The scheme that achieves capacity in very  strong  interference is obtained by setting  $U_{1c}=X_1$ so that both transmitters send only public messages.
The scheme  that we use to show the achievability in the ``strong cognitive decoding" regime mixes these two schemes  by splitting the cognitive message into public and private messages.
This relaxes the strong interference achievability conditions as now the cognitive encoder needs to decode only part of the cognitive message.
The scheme also relaxes the  very weak achievability condition since it allows the cognitive encoder to decode part of the cognitive message and remove its unwanted effects.
For this reason, the resulting achievability conditions are looser than both cases.

\end{rem}
%

\section{Capacity for the semi-deterministic CIFC}
\label{sec:semi-det CIFC}


Consider the specific class of DM-CIFC for which the signal received at receiver~1 is a deterministic function of the channel inputs, that is
\ea{
Y_1=f_1(X_1,X_2).
\label{eq: semi-det equation}
}
This class of channels is termed semi-deterministic CIFC and it was first introduced in \cite{biao2009Arxiv}.
In  \cite{biao2009Arxiv} the capacity region is derived for the case $I(Y_1;X_2)\geq I(Y_2; X_2)$; we extend this result by determining the capacity region in general (no extra conditions). Note that the authors of  \cite{biao2009Arxiv} consider the case where $f_1$ is invertible; we do not require this condition.


\begin{thm}
The capacity region of the semi-deterministic cognitive interference channel such that \reff{eq: semi-det equation} consists of all $(R_1,R_2) \in {\mathbb R}_+^2$ such that
\eas{
R_1 &\leq& H(Y_1|X_2) \\
R_2 &\leq& I(Y_2; U, X_2) \\
R_1+R_2 &\leq& I(Y_2; U,X_2)+H(Y_1| U, X_2)
}{\label{eq: capacity semi-det}}
taken over the union of all distributions $p_{U,X_1,X_2}$.
\label{thm: capacity semideterministic}
\end{thm}

\begin{proof}
\emph{Outer bound:}
The outer bound is obtained from Theorem \ref{thm:outer bound wu} ``one auxiliary RV outer bound" , by using the deterministic condition in \reff{eq: semi-det equation}.

\emph{Achievability:}
Consider the scheme with only the RVs $X_2$, $U_{1pb}$ and $U_{2pb}$, obtained by setting $U_{2c}=U_{1c}=\emptyset$.
The achievable rate region of Theorem \ref{thm:our achievable region} becomes:
\begin{subequations}
\ea{
R'_{1pb}             &\geq& I(U_{1pb}; X_2)
\\
R'_{1pb}+R'_{2pb} &\geq&  I(U_{1pb};U_{2pb}, X_2 )
\\
R_{2pa}+R_{2pb}+R'_{2pb} &\leq& I(Y_2; U_{2pb},X_{2})
\\
R_{2pb}+R'_{2pb} &\leq& I(Y_2; U_{2pb}|X_{2})
\\
R_{1pb}+R'_{1pb} &\leq& I(Y_1; U_{1pb}),
}
\label{eq:simple achievable scheme}
\end{subequations}
taken over the union of all input distributions $p_{U_{1pb},U_{2pb},X_1,X_2} p_{Y_1,Y_2| X_1, X_2}$.

From the Fourier Motzkin elimination of this sub-scheme, we see that we can set $R_{2pb}=0$ without loss of generality and
that the region can be rewritten as
\eas{
\Rc_{0}(U_{1pb},U_{2pb},X_2)\defeq
\{
&R_1     &\leq I(Y_1; U_{1pb})-I(U_{1pb}; X_2) \label{R1: r1}\\
&R_1     &\leq I(Y_2; U_{2pb}|X_2)-I(U_{1pb};U_{2pb}|X_2)+I(Y_1; U_{1pb})-I(U_{1pb};X_2) \label{R1: r1bis}\\
&R_2     &\leq I(Y_2; U_{2pb},X_2) \label{R1: r2}\\
&R_1+R_2 &\leq I(Y_2; U_{2pb},X_2)+I(Y_1; U_{1pb})-I(U_{1pb};U_{2pb}, X_2)\label{R1: r1+r2} \}
}{\label{eq:FME simple scheme 1}}
taken over the union of all distributions that factor as
\ea{
p_{U_{1pb},U_{2pb},X_1,X_2} p_{Y_1,Y_2|X_1,X_2}.
\label{eq: factorization simple scheme}
}
Let
\eas{\Rc_{1}(U_{1pb},U_{2pb},X_2)\defeq\{
&R_1     &\leq I(Y_1; U_{1pb})-I(U_{1pb}; X_2) \label{R0: r1}\\
&R_2     &\leq I(Y_2; U_{2pb},X_2) \label{R0: r2}\\
&R_1+R_2 &\leq I(Y_2; U_{2pb},X_2)+I(Y_1; U_{1pb})-I(U_{1pb};U_{2pb}, X_2)\label{R0: r1+r2} \}
}{\label{eq:FME simple scheme 2}}
and
\eas{
\Rc_{2}(U_{1pb},X_2) \defeq\{
&R_1     &\leq I(Y_1; U_{1pb})-I(U_{1pb}; X_2) \label{R2: r1}\\
&R_2     &\leq I(Y_2; X_2) \label{R2: r2} \}.
}{\label{eq:FME simple scheme 3}}

Notice that
\[
\Rc_{2}(U_{1pb},X_2) \subseteq \Rc_{1}(U_{1pb},U_{2pb},X_2) \subseteq \Rc_{0}(U_{1pb},U_{2pb},X_2),
\]
since
\[
\Rc_{2}(U_{1pb},X_2)=\Rc_{1}(U_{1pb},U_{2pb}=X_2,X_2)=\Rc_{0}(U_{1pb},U_{2pb}=X_2,X_2),
\]
and $\Rc_{0}(U_{1pb},U_{2pb},X_2)$ has one less constraint than $\Rc_{1}(U_{1pb},U_{2pb},X_2)$.

We now wish to show that
\pp{
\bigcup_{p_{X_2,U_{1pb},U_{2pb}}} \Rc_{0}=\bigcup_{p_{X_2,U_{1pb},U_{2pb}}} \Rc_{1},
}
that is, equation \reff{R1: r1bis} can be removed from the Fourier Motzkin eliminated region of \reff{eq:simple achievable scheme}.
The proof of this equivalence follows that of \cite[Lemma 2]{chong2008han}.
For   $P(U_{1pb},U_{2pb},X_2)$ such that
$$
I(Y_2; U_{2pb}|X_2)-I(U_{1pb};U_{2pb}| X_2) \geq 0
$$
we have
$$
\Rc_{1}(U_{1pb},U_{2pb},X_2) = \Rc_{0}(U_{1pb},U_{2pb},X_2).
$$
For those  $P(U_{1pb},U_{2pb},X_2)$ such that
$$
I(Y_2; U_{2pb}|X_2)-I(U_{1pb};U_{2pb}| X_2) < 0
$$
we have that  the point
$$
(R_1,R_2)=(I(Y_1; U_{1pb})-I(U_{1pb};X_2)), I(Y_2; X_2))
$$
is achievable in $\Rcal_2$. This point lies inside $\Rc_{1}$ and $\Rc_{0}$  and satisfies
%
all the rate constraints in \reff{eq:FME simple scheme 1} but \reff{R1: r1bis}.
In particular, the sum rate equation  \reff{R1: r1+r2} given by
\[
R_1+R_2 \leq I(Y_2; U_{2pb},X_2)+I(Y_1; U_{1pb})-I(U_{1pb};U_{2pb}, X_2),
\]
which implies
\[
R_2 \leq I(Y_2; X_2)
\]
since
\pp{
R_2 &\leq I(Y_2; U_{2pb},X_2)+I(Y_1; U_{1pb})-I(U_{1pb};U_{2pb}, X_2)-R_1\\
&=I(Y_2;X_2)+I(Y_2; U_{2pb}|X_2) - I(U_{1pb};U_{2pb}| X_2)\\
&\leq I(Y_2;X_2).
}
Using time sharing we can show the achievability of the whole region $\Rcal_1 \cap \Rcal_0 $, which
means that the rate points that are not in $\Rc_{0}(U_{1pb},U_{2pb},X_2)$
are in $\Rc_{2}(U_{1pb},X_2)$. But since $\Rc_{2}(U_{1pb},X_2)$ is special case
of $\Rc_{0}(U_{1pb},U_{2pb},X_2)$, we conclude that
\[
\Rc_{1}(U_{1pb},U_{2pb},X_2) = \Rc_{0}(U_{1pb},U_{2pb},X_2).
\]
This means is that decoder~2 must not
decode $U_{2pb}$ if that imposes a more stringent rate constraint than
the decoding of $U_{1pb}$ at the intended decoder~1. For this reason $U_{2pb}$ can be chosen so that $U_{2pb}=X_2$ without loss of generality. 
This shows that $\Rcal_1$ is achievable and thus concludes the achievability proof.

\end{proof}
\begin{rem}
The achievable scheme of equation \reff{eq:simple achievable scheme} cannot be obtained as a special case of any previously known achievable scheme but \cite{jiang-achievable-BCCR}.
The RV $U_{2pb}$, which broadcasts the private primary message  from transmitter~1, appears in \cite{biao2009Arxiv} as well.
In this scheme though is not possible to reobtain the scheme of equation \reff{eq:simple achievable scheme} with a specific choice of the RVs since
 the same message $w_{2pa}$ is transmitted in $U_{2pb}$  and the private primary message $X_2$. 
\end{rem}
\section{Capacity for the deterministic CIFC}
\label{sec:det CIFC}

In the  deterministic CIFC both outputs are deterministic functions of the channel inputs, that is
\ea{
Y_1= Y_1 (X_1,X_2) \nonumber \\
Y_2= Y_2 (X_1,X_2). \label{eq: det equation}
}
This class of channels is a subclass  of the semi-deterministic CIFC of Section \ref{sec:semi-det CIFC}, and we already have obtained the capacity region for this case. However, we re-derive the capacity region in a new fashion for this channel, which illustrates two new ideas:

1) We show the achievability of the outer bound of Theorem \ref{thm: outer bound CIFC} when letting $Y_2'=Y_2$, instead of the  outer bound of Theorem \ref{thm:outer bound wu} ``one auxiliary RV outer bound".

2) We show achievability of this outer bound using a single unified scheme. 

\begin{thm}
The capacity region of the deterministic cognitive interference channel consists of all $(R_1,R_2) \in {\mathbb R}_+^2$ such that
\eas{
R_1 &\leq& H(Y_1|X_2) \\
R_2 &\leq& H(Y_2) \\
R_1+R_2 &\leq& H(Y_2)+H(Y_1| Y_2, X_2)
}{\label{eq: capacity det}}
taken over the union of all distributions $p_{X_1,X_2}$.
\label{thm: capacity deterministic}
\end{thm}

\begin{proof}
\emph{Outer bound:}
The outer bound is obtained from Theorem \ref{thm: outer bound CIFC} using the deterministic conditions in \reff{eq: det equation}.

\emph{Achievability:}
Consider the scheme  in \reff{eq:FME simple scheme 2} and let $U_{1pb}=Y_1$, $U_{2pb}=Y_2$ to achieve the region
\eas{
R_1 &\leq& H(Y_1|X_2)\\
R_2 &\leq& H(Y_2)\\
R_1+R_2 &\leq& H(Y_2; U,X_2)+H(Y_1|Y_2, X_2)
}{}
which corresponds to the outer bound in \reff{eq: capacity det}. 
\end{proof}

\section{Examples}
\label{sec:examples}
The scheme that achieves capacity in the deterministic and semi-deterministic  CIFC uses the RV $U_{2pb}$ to perform Gel'fand Pinsker binning  to achieve the most general distribution among $(X_2,U_{1pb}, U_{2pb})$, but interestingly,  carries no message.
This feature of the capacity achieving scheme does not provide a clear intuition on the role of the RV $U_{2pb}$.
For this reason we present  two examples of deterministic channels where the encoders can choose their respective codebooks in a way that allows binning of the interference without rate splitting.
%
%
To make these examples more interesting we choose them so that they do {\it not} fall into the category of the ``very strong interference regime" of Theorem \ref{thm: Very strong interference capacity} that in the deterministic case reduces to
\ea{
H(Y_1|X_2) &\leq& H(Y_2|X_2) \nonumber \\
H(Y_2) &\leq& H(Y_1) \ \ \lag \forall p_{X_1,X_2}.
\label{eq: det storng interference conditions}
}
Unfortunately, checking for the ``very weak interference condition" of Theorem \ref{thm:Very weak interference capacity} is not possible as no cardinality bounds on $U$ are available.

\subsection{Example I: the ``Asymmetric Clipper"}
\label{secIV:example I}

\begin{figure}
\begin{center}
  \includegraphics[width=8cm]{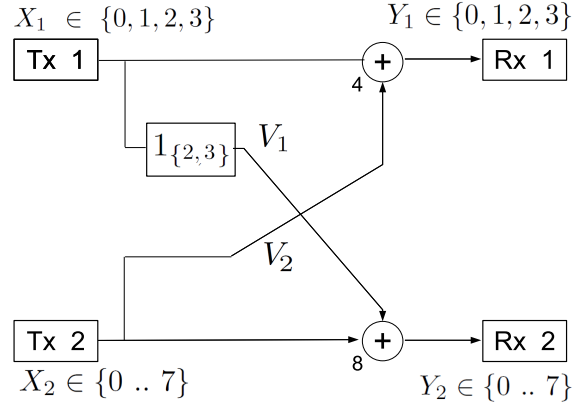}\\
  \caption{The ``asymmetric clipper" of Section \ref{secIV:example I}.}
  \label{example I}
\end{center}
\end{figure}

Consider the channel in Fig.~\ref{example I}.
The input and output alphabets are  $\Xcal_1 = \Ycal_1 = \{0,1,2,3\}$
and $\Xcal_2 = \Ycal_2 =  \{0,1,2,3,4,5,6,7 \}$ and
the input/output relationships are
\begin{align}
Y_1 &= X_1 \oplus_4  X_2, \\
Y_2 &= 1_{\{2,3\}}(X_1)   \oplus_8 +X_2,
\label{eq:channel example I}
\end{align}
where $1_{A}(x)=1$ if $x\in A$ and zero otherwise,  and $\oplus_N$ denotes the addition operation over the Galois field $\Gcal_N$ defined as the modulo sum over elements in the  finite field  $\{1...N\}$.
Also let $\Ucal(\Sc)$ be the uniform distribution over the set $\Sc$.

First we show  that the channel in~\reff{eq:channel example I} does not fall in the ``very strong interference" class.

Consider the input distribution:
\pp{
X_2 \sim \Ucal(1) \ \implies \ P[X_1=0]=1, \\
X_2 \sim  \Ucal( \Xcal_2).
}
For this input distribution, we have
$Y_1 \sim \Ucal( \Ycal_1) $  and $Y_2 \sim \Ucal( \Ycal_2)$, so that
$$
H(Y_2)=\log(| \Ycal_2|)=3>2=\log(| \Ycal_1|)=H(Y_1)
$$
which does not satisfy the ``very strong interference'' condition of \reff{eq: det storng interference conditions}. 
%

For this  channel we have:
\begin{align*}
 H(Y_1|X_2) &\leq H(Y_1) \leq \log(| \Ycal_1|)=2 \\
 H(Y_2)     &\leq \log(| \Ycal_2|)=3 \\
 H(Y_1|X_2 ,Y_2) &\leq H(X_1|1_{\{2,3\}}(X_1)) \leq 1.
 \end{align*}
where the last bound follows from  the multiplicity of the solutions of an addition in a Galois field.
This shows that the outer bound in Theorem \ref{thm: capacity deterministic} is included in
\begin{subequations}
\ea{
R_1 & \leq  2 \\
R_2 & \leq  3 \\
R_1+R_2 & \leq 4.
}
\label{eq: outer bound example}
\end{subequations}
We now show that the region in \reff{eq: outer bound example} indeed corresponds to the  Theorem \ref{thm: capacity deterministic} when considering the union over all input distributions.
The corner point $(R_1,R_2)=(1,3)$ in~\reff{eq: outer bound example} is obtained in Theorem~\ref{thm: capacity deterministic}
with the input distribution:
\pp{
X_1 \sim \Ucal(\{0,1\}) \\
X_2 \sim \Ucal(\Xcal_2).
}
The corner point $(R_1,R_2)=(2,2)$ in~\reff{eq: outer bound example} is obtained in Theorem~\ref{thm: capacity deterministic} by considering the input distribution:
\pp{
X_1 \sim \Ucal(\Xcal_1) \\
X_2 \sim \Ucal(\Xcal_2).
}
Time sharing shows that the region of ~\reff{eq: outer bound example} and the region of Theorem ~\ref{thm: capacity deterministic}  indeed coincide.

We next show the achievability of the corner point $(R_1,R_2)=(1,3)$.  Consider the following strategy:
\begin{itemize}
  \item transmitter~2 sends symbols from $\Xcal_2=\{0...7\}$ with uniform probability,
  \item transmitter~1 transmits $\lsb x_1-x_2\rsb_2$ (where the inverse of the difference operation is taken over the ring $\Gcal_2$);
  \item receiver~1 decodes $\wh_1=\lfloor \f {y_2} 2 \rfloor$;
  \item receiver~2 decodes $\wh_2=y_2 $.
\end{itemize}

It can be verified by inspection of Table \ref{tab:13} that the rate pair $(R_1,R_2)=(1,3)$
is indeed achievable.

\begin{table} [h!h!h!h!h!h!h!h!h!]
\center
\caption {Achievability for $(R_1,R_2)=(1,3)$ in Example I in Section \ref{secIV:example I}.}
\begin{tabular}{l|llllllll}
\hline
&$\om_1$ & $\om_2$ & $x_1$ & $x_2$ & $y_1$ & $y_2$ & $\omh_1$ & $\omh_2$ \\
\hline
0 & 0 & 0 & 0 & 0 & 0 & 0 & 0 & 0 \\
1 & 0 & 1 & 1 & 0 & 2 & 2 & 1 & 0 \\
2 & 0 & 2 & 0 & 2 & 2 & 2 & 2 & 0 \\
3 & 0 & 3 & 1 & 3 & 2 & 0 & 3 & 0 \\
4 & 0 & 4 & 0 & 4 & 0 & 0 & 4 & 0 \\
5 & 0 & 5 & 1 & 5 & 0 & 2 & 5 & 0 \\
6 & 0 & 6 & 0 & 6 & 0 & 2 & 6 & 0 \\
7 & 0 & 7 & 1 & 7 & 0 & 0 & 7 & 0 \\
\hline
8  & 1 & 0 & 1 & 0 & 0 & 1 & 0 & 1 \\
9  & 1 & 1 & 0 & 0 & 2 & 1 & 1 & 1 \\
10 & 1 & 2 & 1 & 2 & 2 & 3 & 2 & 1 \\
11 & 1 & 3 & 0 & 3 & 2 & 3 & 3 & 1 \\
12 & 1 & 4 & 1 & 4 & 0 & 1 & 4 & 1 \\
13 & 1 & 5 & 0 & 5 & 0 & 1 & 5 & 1 \\
14 & 1 & 6 & 1 & 6 & 0 & 3 & 6 & 1 \\
15 & 1 & 7 & 0 & 7 & 0 & 3 & 7 & 1 \\\hline
\end{tabular}
\label{tab:13}
\end{table}

Now we  show the achievability of the corner point $(R_1,R_2)=(2,2)$.  Consider the following strategy:
\begin{itemize}
  \item transmitter~2 sends symbols from $x_2 \in \{0,2,4,6\}$ with uniform probability;
  \item transmitter~1 transmits $\lsb x_1-x_2\rsb_4$ (where the inverse of the difference operation is taken over the ring $\Gcal_4$);
  \item receiver~1 decodes $\wh_1=y_1$;
  \item receiver~2 decodes $\wh_2=\lfloor \f {y_2} 2 \rfloor $.
\end{itemize}
It can be verified by the inspection of Table \ref{tab:22} that the rate pair $(R_1,R_2)=(2,2)$
is indeed achievable.

\begin{table} [h!h!h!h!h!h!h!h!h!]
\center
\caption{Achievability table for the rate point $(R_1,R_2)=(2,2)$ in Example I in Section \ref{secIV:example I}.}
\label{tab:22}
\begin{tabular}{l|llllllll}
\hline
& $\om_1$ & $\om_2$ & $x_1$ & $x_2$ & $y_1$ & $y_2$ & $\omh_1$ & $\omh_2$ \\
\hline
0 &0 & 0 & 0 & 0 & 0 & 0 & 0 & 0 \\
1 &0 & 1 & 2 & 2 & 0 & 3 & 0 & 1 \\
2 &0 & 2 & 0 & 4 & 0 & 4 & 0 & 2 \\
3 &0 & 3 & 2 & 6 & 0 & 7 & 0 & 3 \\
\hline
4 & 1 & 0 & 1 & 0 & 1 & 0 & 1 & 0 \\
5 &1 & 1 & 3 & 2 & 1 & 3 & 1 & 1 \\
6 &1 & 2 & 1 & 4 & 1 & 4 & 1 & 2 \\
7 &1 & 3 & 3 & 6 & 1 & 7 & 1 & 3 \\
\hline
8 &2 & 0 & 2 & 0 & 2 & 0 & 2 & 0 \\
9  &2 & 1 & 0 & 2 & 2 & 2 & 2 & 1 \\
10 &2 & 2 & 2 & 4 & 2 & 5 & 2 & 2 \\
11 &2 & 3 & 0 & 6 & 2 & 6 & 2 & 3 \\
\hline
12 & 3 & 0 & 3 & 0 & 3 & 1 & 3 & 0 \\
13 &3 & 1 & 1 & 2 & 3 & 2 & 3 & 1 \\
14 &3 & 2 & 3 & 4 & 3 & 5 & 3 & 2 \\
15 &3 & 3 & 1 & 6 & 3 & 6 & 3 & 3 \\
\hline
\end{tabular}
\end{table}


In this example we see how the two senders jointly design the codebook to achieve the outer bound and in particular how the cognitive transmitter~1 adapts its strategy to the transmissions from the primary pair so as to avoid interfering with it.

In achieving the point $(R_1,R_2)=(1,3)$, transmitter~2's strategy is that of  a point to point channel. 
Transmitter~1 chooses its codewords so as not to  interfere with the primary transmission. Only two codewords do not interfere: it alternatively picks one of these two codewords to produce the desired channel output. For example, when the primary message is sending $\om_2=0$ (line $0$ and $8$ in Table \ref{tab:13}) transmitter~1  can send either $1$ or $2$ without creating interference at receiver~2. On the other hand, these two values produce a different output at receiver~1, allowing the transmission of 1~bit.

In achieving the point $(R_1,R_2)=(2,2)$, the primary receiver picks its codewords so as to tolerate 1~unit of interference.
Transmitter~1 again chooses its input codewords in order to create at most 1~unit of interference at the primary decoder.
By adapting its transmission to the primary symbol, the cognitive transmitter is able to always find four such codewords.
It is interesting to notice the tension at transmitter~1 between the interference it creates at the primary decoder and its own rate.
There is an optimal trade off between these two quantities that is achieved by carefully picking the codewords at the primary transmitter.
For example, when the primary receiver is sending $\om_2=0$ (lines $0,4,8$ and $12$), transmitter~1 can send $x_1 \in \{0,1,2,3\}$ and create at most 1~bit of interference at receiver~2. Each of these four values produces  a different output at receiver~1], thus allowing the transmission of 2~bits.

\subsection{Example II: the ``Symmetric Clipper"}
\label{sec:example II}

\begin{figure}
\begin{center}
  \includegraphics[width=8cm]{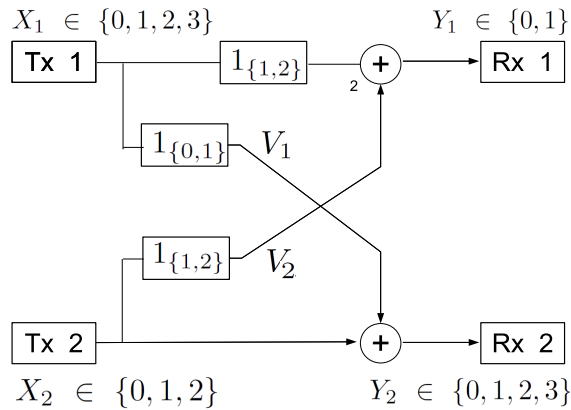}\\
  \caption{``Symmetric Clipper" of Section \ref{sec:example II}}\label{exampleII}
\end{center}
\end{figure}

Consider the now channel in Fig. \ref{exampleII}.
The channel input and output alphabets  are  $\Xcal_1 = \{0,1,2,3\} = \Ycal_2$,
$\Xcal_2 \in \{0,1,2\}$, and $\Ycal_1 = \{0,1\}$.
The input/output relationships are:
\pp{
Y_1=1_{\{1,2\}}(X_1) \oplus_2  1_{\{1,2\}}(X_2)\\
Y_2=1_{\{0,1\}}(X_1)   \oplus X_2
}
Consider the input distribution:
%
%
%
Consider the input distribution:
\pp{
P[X_1 =3]=1,\\
X_2 \sim \Ucal (\{1,2\}),
}
in this case $H(Y_1)=0$ and $H(Y_2)=1$. 
This shows that there exists at least one input distribution for which $H(Y_2)> H(Y_1)$ and thus this channel is not in the ``very strong interference" regime.
The outer bound of Theorem~\ref{thm: capacity deterministic}  is achieved here by a single input distribution $p_{X_1,X_2}$: consider the distribution in Table  \ref{tab:exII}. This distribution produces $H(Y_1)=1=\log_2(|\Ycal_1|)$ and $H(Y_2)=2=\log( |\Ycal_2|)$  and clearly no larger outer bound can exist given the output cardinality. We therefore conclude that the region of  Theorem~\ref{thm: capacity deterministic}  can be rewritten as:
\pp{
R_1 \leq 1 \\
R_2 \leq 2.
}
\begin{table}
  \centering
  \caption{The input distribution for Example II }\label{tab:exII}
\begin{tabular}{|l|l|l|l|l|l|}
\cline{2-5}
\multicolumn{1}{l|}{$X_2$ \ $X_1$} & \multicolumn{1}{l}{1} & \multicolumn{1}{l}{2} & \multicolumn{1}{l}{3} & 4 & \multicolumn{1}{l}{} \\
\hline
0 & 1/8 & 1/8 & 1/8 & 1/8 & 1/2 \\
\cline{2-5}
1 & 1/8 & 1/8 & 0 & 0 & 1/4 \\
\cline{2-5}
2 & 1/8 & 1/8 & 0 & 0 & 1/4 \\
\hline
\multicolumn{1}{l|}{} & \multicolumn{1}{l}{3/8} & \multicolumn{1}{l}{3/8} & \multicolumn{1}{l}{1/8} & 1/8 & \multicolumn{1}{l}{} \\
\cline{2-5}
\end{tabular}
\end{table}

This region can be shown to be achievable using the transmission scheme  described in Table \ref{Example II}.
\begin{table}
\centering
\caption{Achievability table for the rate point $(R_1,R_2)=(1,2)$ in Example II.}\label{Example II}
\begin{tabular}{l|llllllll}
\hline
 &$\om_1$ & $\om_2$ & $x_1$ & $x_2$ & $v_1$ & $v_2$ & $y_1$ & $y_2$ \\
\hline
0 &0 & 0 & 3 & 0 & 0 & 0 & 0 & 0 \\
1 &0 & 1 & 0 & 0 & 1 & 0 & 0 & 1 \\
2 &0 & 2 & 1 & 1 & 1 & 1 & 0 & 2 \\
3 &0 & 3 & 1 & 2 & 1 & 1 & 0 & 3 \\
\hline
4 &1 & 0 & 2 & 0 & 0 & 0 & 1 & 0 \\
5 &1 & 1 & 1 & 0 & 1 & 0 & 1 & 1 \\
6 &1 & 2 & 0 & 1 & 1 & 1 & 1 & 2 \\
7 &1 & 3 & 0 & 2 & 1 & 1 & 1 & 3 \\
\hline
\end{tabular}
\end{table}
The decoding is simply
$\omh_i=Y_i, \ i\in\{1,2\}.$
This transmission scheme achieves the proposed outer bound, thus showing capacity.
The transmission scheme can be described as follows:
\begin{itemize}
  \item  encoder 2 transmits $[x_2-1]^+$;
  \item  encoder 1 transmits the value $X_1$ that simultaneously makes $Y_1=\om_1$ and $Y_2=\om_2$. For each $\om_1$ and $\om_2$ such a value always exists because $X_2$ takes on only three possible values;
  \item  receivers~1 and ~2 decode $\omh_1=Y_1$ and $\omh_2=Y_2$.
 \end{itemize}

This example is particularly interesting since both decoders obtain the transmitted symbol without suffering any interference from the other user.
Here cognition allows the simultaneous cancelation of the interference at both decoders.
Encoder~2 has only three codewords and relies on transmitter~1 to achieve its full rate of $R_2=2$. In fact encoder~1 is able to design its codebook to transmit two codewords for its decoder and still contribute to the rate of primary user by making the codewords corresponding to $\om_2=\{2,3\}$ distinguishable at the cognitive decoder.

This feature of the capacity achieving scheme is intriguing: the primary transmitter needs the support of the cognitive transmitter to achieve $R_2=2$ since its input alphabet has cardinality three. The transmitters optimally design their codebooks so to make the  effect  $X_1$ on both outputs the desired one.

For example consider the transmission of $\om_2=2$ or $3$ (lines $2,3,6$ and $7$). In this case transmitter~1 sends $x_1=0$ or  $x_1=1$ to simultaneously influence both channel outputs so that both decoders receive the desired symbols. This simultaneous cancelation is possible due to  the channel's deterministic nature and the extra message knowledge at the cognitive transmitter.

\section{Conclusion}
\label{Conclusion}

In this paper we focused on the  discrete memoryless cognitive interference channel and derived new inner and outer bounds,  derived the capacity region for a class of ``better cognitive decoding" channels, and obtained the capacity region for the semi-deterministic cognitive interference channel.
%
We proposed a new outer bound 
 using an idea originally devised for the broadcast channel in \cite{sato1978outer}.
This outer bound does not involve auxiliary RVs and is thus more easily computable. 
Our outer bound is in general looser than the outer bound in \cite{WuDegradedMessageSet} and they coincide
in the ``strong interference" regime of \cite{Maric_strong_interference}.
We also proposed a new inner bound that generalizes all other known achievable rate regions.
In particular we showed the inclusion of the region of \cite{devroye2005cognitive, DevroyeThesis};  it was previously unclear how the performance of the scheme in \cite{devroye2005cognitive, DevroyeThesis} compared with that of other achievable rate regions. %
We determined capacity for a class of channels that we term
the ``better cognitive decoding" regime. The conditions defining this regime are  looser than the ``very weak interference condition" of \cite{WuDegradedMessageSet} and the ``very strong interference condition" of \cite{Maric_strong_interference}  and is the largest region where capacity is known.
We also determined the capacity region for the class of semi-deterministic cognitive interference channels where the output at the cognitive receiver is a deterministic function of the channel inputs.
Furthermore, for channels where  both outputs are deterministic functions of the inputs, we  showed the achievability of our new outer bound.
This result shows that our outer bound, even though looser than the outer bound in \cite{WuDegradedMessageSet}, is tight for certain channels.
The scheme that achieves capacity in the deterministic cognitive interference channel uses Gelf'and-Pinsker binning against the interference created at the primary receiver. This binning is  performed by the cognitive encoder for the cognitive decoder.
This feature of the transmission scheme was never known before to be capacity achieving.
We conclude the paper by presenting two examples that show new interesting features of the capacity achieving scheme in the deterministic cognitive interference channel.
%
%
%
%
Extensions of the results presented here to Gaussian channels will be presented in \cite{RTDjournal2}.

\appendix

\subsection{Error analysis of the achievable region $\Rcal_{RTD}$ of Theorem \ref{thm:our achievable region} }
\label{app:error analysis RTD}
Without loss of generality assume that the message
$(w_{1c},w_{2c},w_{2pa},w_{1pb},w_{2pb})=(1,1,1,1,1)$
was sent and let $(\bo_0,\bo_1,\bo_2)$ be the tuple $(b_0,b_1,b_2)$
chosen at encoder~1.
Let  $(\wh_{1c},\wh_{2c},\wh_{2pa},\wh_{2pb},\hat{b}_0,\hat{b}_2)$
be the estimate at the decoder~2 and
$(\widehat{\wh}_{1c},\widehat{\wh}_{2c},\widehat{\wh}_{1pb},\hat{\hat{b}}_0,\hat{\hat{b}}_1)$
be the estimate at the decoder~1.

The probability of error at decoder~$u$, $u\in\{1,2\}$, is bounded by
\[
P[{\rm error}~u]
\leq
 P[{\rm error}~u| {\rm encoding\,successful}]
+P[{\rm encoding\,NOT\,successful}].
\]
An encoding error occurs if encoder~1 is not able to find
a tuple $(\bo_0,\bo_1,\bo_2)$ that guarantees typicality.
A decoding error is committed at decoder 1 when $(\widehat{\wh}_{1c},\widehat{\wh}_{1pb})\neq (1,1)$.
A decoding error is committed at decoder 2 when $(\wh_{2c},\wh_{2pa},\wh_{2pb})\neq (1,1,1)$.

\subsection{Encoding Error}
The probability that the encoding fails can be bounded as:
\pp{
P[{\rm encoding\,NOT\,successful}]=P \lsb
\bigcap_{b_0=1}^{2^{N R'_{1c}  }} \bigcap_{b_1=1}^{2^{N R'_{1pb} }}  \bigcap_{b_2=1}^{2^{N R'_{2pb}} } \rnone \\
\lag \lnone \lb U_{2c}^N(1), X_{2}^N(1,1),U_{1c}^N(1,1,b_0),U_{1pb}^N(1,1,b_0, 1,b_1), U_{2pb}^N(1,1,b_0, 1,b_2) \rb  \notin T_\epsilon^N (p_{U_{2c},X_{2},U_{1c},U_{1pb},U_{2pb}})
\rsb \\
\lag \lag= P[K=0] \leq \f{\var[K]}{E^2[K]}
}
where
\[
K=\sum_{b_0=1}^{2^{N R'_{1c}  }} \sum_{b_1=1}^{2^{N R'_{1pb} }}  \sum_{b_2=1}^{2^{N R'_{2pb}} }  K_{b_0,b_1,b_2}
\]
and
\[
K_{b_0,b_1,b_2}= 1_{ \lcb \lb U_{2c}^N(1), X_{2}^N(1,1),U_{1c}^N(1,1,b_0),U_{1pb}^N(1,1,b_0, 1,b_1), U_{2pb}^N(1,1,b_0, 1,b_2) \rb
\in T_\epsilon^N (p_{U_{2c},X_{2},U_{1c},U_{1pb},U_{2pb}}) \rcb},
\]
where $1_{\{x\in A\}}=1$ if $x\in A$ and zero otherwise.

The mean value of $K$ (neglecting all terms that depend on $\ep$ and that eventually go to zero) is:
\begin{align*}
E[K]
= \sum_{b_0=1}^{2^{N R'_{1c}  }} \sum_{b_1=1}^{2^{N R'_{1pb} }}  \sum_{b_2=1}^{2^{N R'_{2pb}} } P[K_{b_0,b_1,b_2}=1]
= 2^{N(R'_{1c}  +R'_{1pb} +R'_{2pb}-A)}
\end{align*}
with
\begin{align*}
2^{-N A}
  &=P[K_{b_0,b_1,b_2}=1]=E[K_{b_0,b_1,b_2}]
\\&=P[ \lb U_{2c}^N(1), X_{2}^N(1,1),U_{1c}^N(1,1,b_0),U_{1pb}^N(1,1,b_0, 1,b_1), U_{2pb}^N(1,1,b_0, 1,b_2) \rb
\in T_\epsilon^N (p_{U_{2c},X_{2},U_{1c},U_{1pb},U_{2pb}}) ]
\\
&=\sum_{ (u^N_{1c},u^N_{1pb},u^N_{2pb})
\in T_\epsilon^N (p_{U_{2c},X_{2},U_{1c},U_{1pb},U_{2pb}}|u^N_{2c},x^N_{2})}
p_{U_{1c}|U_{2c}}\, p_{U_{2pb}|U_{2c},U_{1c},X_{2}}\,p_{U_{1pb}|U_{2c},U_{1c}}
\\
&\geq
2^{-N [I(U_{1c}; X_{2} | U_{2c})+I(U_{1pb}; X_{2}, U_{2pb}| U_{1c}, U_{2c})]}.
\end{align*}

The variance of $K$ (neglecting all terms that depend on $\ep$ and that eventually go to zero) is:
\begin{align*}
\var[K]
& = \sum_{b_0=1}^{2^{N R'_{1c}  }}\sum_{b_1=1}^{2^{N R'_{1pb} }} \sum_{b_2=1}^{2^{N R'_{2pb}}}
    \sum_{b_0'=1}^{2^{N R'_{1c}  }}\sum_{b_1'=1}^{2^{N R'_{1pb} }} \sum_{b_2'=1}^{2^{N R'_{2pb}}}
\lb P[K_{b_0,b_1,b_2}=1,K_{b'_0,b'_1,b'_2}=1]-P[K_{b_0,b_1,b_2}=1]P[K_{b'_0,b'_1,b'_2}=1]\rb
\\&=\sum_{b_0'=b_0,(b_1,b_2,b_1',b_2')}
\lb P[K_{b_0,b_1,b_2}=1,K_{b_0,b_1',b_2'}=1]-P[K_{b_0,b_1,b_2}=1]P[K_{b_0,b_1',b_2'}=1]\rb
\\&\leq \sum_{b_0,(b_1,b_2,b_1',b_2')}
   P[K_{b_0,b_1,b_2}=1,K_{b_0,b_1',b_2'}=1]
\end{align*}
because when $b_0\neq b_0'$, that is, $U^N_{1c}(...,b_0)$ and $U^N_{1c}(...,b'_0)$ are independent,
the RVs $K_{b_0,b_1,b_2}$ and $K_{b'_0,b_1',b_2'}$ are independent
and they do not contribute to the summation.
We thus can focus only on the case $b_0=b_0'$.  We can write:
\begin{align*}
\var[K]
&\leq \underbrace{\sum_{b_0, \,b_1=b_1',\, b_2=b_2'}
 P[K_{b_0,b_1,b_2}=1]}_{=E[K]}
\\&+ \underbrace{\sum_{b_0, \,b_1=b_1',\, b_2\not=b_2'}
 P[K_{b_0,b_1,b_2}=1]P[K_{b_0,b_1,b_2'}=1|K_{b_0,b_1,b_2}=1]}_{=E[K]\, 2^{N (R'_{2pb}-B)}}
\\&+ \underbrace{\sum_{b_0, \,b_1\not=b_1',\, b_2=b_2'}
 P[K_{b_0,b_1,b_2}=1]P[K_{b_0,b_1',b_2}=1|K_{b_0,b_1,b_2}=1]}_{=E[K]\, 2^{N (R'_{1pb} -C)}}
\\&+ \underbrace{\sum_{b_0,\,b_1\not=b_1',\, b_2\not=b_2'}
 P[K_{b_0,b_1,b_2}=1]P[K_{b_0,b_1',b_2'}=1|K_{b_0,b_1,b_2}=1]}_{=E[K]\,2^{N (R'_{1pb} +N R'_{2pb}-D)}}
\end{align*}
and
\begin{align*}
2^{-N B}
  &= P[K_{b_0,b_1,b_2'}=1|K_{b_0,b_1,b_2}=1]
\\&= P[  \lb U_{2c}^N(1), X_{2}^N(1,1),U_{1c}^N(1,1,b_0),U_{1pb}^N(1,1,b_0, 1,b_1), U_{2pb}^N(1,1,b_0, 1,b_2') \rb
\in T_\epsilon^N (p_{U_{2c},X_{2},U_{1c},U_{1pb},U_{2pb}}) |
\\&\qquad\lb U_{2c}^N(1), X_{2}^N(1,1),U_{1c}^N(1,1,b_0),U_{1pb}^N(1,1,b_0, 1,b_1), U_{2pb}^N(1,1,b_0, 1,b_2) \rb
\in T_\epsilon^N (p_{U_{2c},X_{2},U_{1c},U_{1pb},U_{2pb}}) ]
\\&= \sum_{ u^N_{2pb}
\in T_\epsilon^N (p_{U_{2c},X_{2},U_{1c},U_{1pb},U_{2pb}}|u^N_{2c},x^N_{2},u^N_{1c},u^N_{1pb})}
p_{U_{2pb}|U_{2c},U_{1c},X_{2}}
\\&=
2^{-N I(U_{2pb}; U_{1pb}|U_{2c},U_{1c},X_{2})},
\end{align*}
and
\begin{align*}
2^{-N C}
  &= P[K_{b_0,b_1',b_2}=1|K_{b_0,b_1,b_2}=1]
\\&= P[  \lb U_{2c}^N(1), X_{2}^N(1,1),U_{1c}^N(1,1,b_0),U_{1pb}^N(1,1,b_0, 1,b_1'), U_{2pb}^N(1,1,b_0, 1,b_2) \rb
\in T_\epsilon^N (p_{U_{2c},X_{2},U_{1c},U_{1pb},U_{2pb}}) |
\\&\qquad\lb U_{2c}^N(1), X_{2}^N(1,1),U_{1c}^N(1,1,b_0),U_{1pb}^N(1,1,b_0, 1,b_1), U_{2pb}^N(1,1,b_0, 1,b_2) \rb
\in T_\epsilon^N (p_{U_{2c},X_{2},U_{1c},U_{1pb},U_{2pb}}) ]
\\&= \sum_{ u^N_{1pb}
\in T_\epsilon^N (p_{U_{2c},X_{2},U_{1c},U_{1pb},U_{2pb}}|u^N_{2c},x^N_{2},u^N_{1c},u^N_{2pb})}
p_{U_{1pb}|U_{2c},U_{1c}}
\\&=
2^{-N I(U_{1pb}; X_{2}, U_{2pb}| U_{1c}, U_{2c})},
\end{align*}
and
\begin{align*}
2^{-N D}
  &=P[K_{b_0,b_1',b_2'}=1|K_{b_0,b_1,b_2}=1]
\\&= P[  \lb U_{2c}^N(1), X_{2}^N(1,1),U_{1c}^N(1,1,b_0),U_{1pb}^N(1,1,b_0, 1,b_1'), U_{2pb}^N(1,1,b_0, 1,b_2') \rb
\in T_\epsilon^N (p_{U_{2c},X_{2},U_{1c},U_{1pb},U_{2pb}}) |
\\&\qquad\lb U_{2c}^N(1), X_{2}^N(1,1),U_{1c}^N(1,1,b_0),U_{1pb}^N(1,1,b_0, 1,b_1), U_{2pb}^N(1,1,b_0, 1,b_2) \rb
\in T_\epsilon^N (p_{U_{2c},X_{2},U_{1c},U_{1pb},U_{2pb}}) ]
\\&= \sum_{ (u^N_{1pb},u^N_{2pb})
\in T_\epsilon^N (p_{U_{2c},X_{2},U_{1c},U_{1pb},U_{2pb}}|u^N_{2c},x^N_{2},u^N_{1c})}
p_{U_{2pb}|U_{2c},U_{1c},X_{2}}\,p_{U_{1pb}|U_{2c},U_{1c}}
\\&=
2^{-N I(U_{1pb}; X_{2}, U_{2pb}| U_{1c}, U_{2c})}
=2^{-N C}.
\end{align*}

Hence, we can bound $P[K=0]$ as:
\pp{
0\leq
P[K=0]
&\leq \displaystyle\frac{
1
+2^{N( R'_{1pb}            -C)}
+2^{N(            R'_{2pb} -B)}
+2^{N( R'_{1pb} +R'_{2pb} -C)}
}{
 2^{N(R'_{1c}  +R'_{1pb} +R'_{2pb}-I(U_{1c}; X_{2} | U_{2c})-C)}
}
}
and $P[K=0]\to 0$ if
\begin{align*}
  &R'_{1c}  +R'_{1pb} +R'_{2pb}-I(U_{1c}; X_{2} | U_{2c})-C                             >0
\\&R'_{1c}  +R'_{1pb} +R'_{2pb}-I(U_{1c}; X_{2} | U_{2c})-C -(            R'_{2pb} -B)>0
\\&R'_{1c}  +R'_{1pb} +R'_{2pb}-I(U_{1c}; X_{2} | U_{2c})-C -( R'_{1pb}            -C)>0
\\&R'_{1c}  +R'_{1pb} +R'_{2pb}-I(U_{1c}; X_{2} | U_{2c})-C -( R'_{1pb} +R'_{2pb} -C)>0
\end{align*}
that is, if
\begin{align*}
  R'_{1c}  +R'_{1pb} +R'_{2pb}&> I(U_{1c}; X_{2} | U_{2c})+ I(U_{1pb}; X_{2}, U_{2pb}| U_{1c}, U_{2c})
                               \\&= I(U_{1c},U_{1pb}; X_{2}|  U_{2c})+I(U_{1pb}; U_{2pb}| U_{1c}, U_{2c},X_{2})
\\R'_{1c}  +R'_{1pb}           &> I(U_{1c}; X_{2} | U_{2c})+ I(U_{1pb}; X_{2}| U_{1c}, U_{2c})
                               \\&= I(U_{1c},U_{1pb}; X_{2}|  U_{2c})
\\R'_{1c}             +R'_{2pb}&> I(U_{1c}; X_{2} | U_{2c}),
\\R'_{1c}                        &> I(U_{1c}; X_{2} | U_{2c})
\end{align*}
as in~\reff{eq:our achievable region R0"}-\reff{eq:our achievable region R1"+R2"}, because the
second to last equation is redundant.


\subsection{Decoding Errors at decoder 2}
\begin{table}
  \centering
  \caption{Error events at decoder 2.}\label{tab:Errors2}
\begin{tabular}{|l|l|l|l|l|l|}
\hline
Event      & $w_{2c}$ & $(w_{1c},b_1)$ & $w_{2pa}$ & $w_{2pb}$ & $p_{Y_2|\star}$ \\\hline
$E_{2,1} $ & X &$\cdots$&$\cdots$&$\cdots$& $p_{Y_2}$ \\ \hline
$E_{2,2a}$ & 1 & X & X &$\cdots$& $p_{Y_2|U_{2c}}$ \\ \hline
$E_{2,2b}$ & 1 & 1 & X &$\cdots$& $p_{Y_2|U_{2c},U_{1c}}$ \\ \hline
$E_{2,3a}$ & 1 & X & 1 & X & $p_{Y_2|U_{2c}       ,X_{2}}$ \\ \hline
$E_{2,3b}$ & 1 & 1 & 1 & X & $P_{Y_2|U_{2c},U_{1c},X_{2}}$ \\\hline
\end{tabular}
\end{table}

If decoder~2 decodes $(\wh_{2c}, \wh_{2pa}, \wh_{2pb})\not=(1,1,1)$, then an error is committed.
The probability of error at decoder~2 is bounded as:
\[
P[{\rm error}~2| {\rm encoding\,successful}]\leq \sum_{i\in\{1,2a,2b,3a,3b\}} P[E_{2,i}],
\]
where $E_{2,i}$, $i\in\{1,2a,2b,3a,3b\}$, are the error events defined in Table \ref{tab:Errors2}.
In Table~\ref{tab:Errors2},
an ``X" means that the corresponding message is in error
(when the header of the column contains two indices,
an ``X" indicates that at least one of the two indexes is wrong),
a ``1" means that the corresponding message is correct,
while the dots ``$\cdots$" indicates that
``{\em it does not matter whether the corresponding message is correct or not;
in this case the most restrictive case is when the message is actually wrong.}"
The last column of Table~\ref{tab:Errors2} specifies the $p_{Y_2|\star}$ to be used in~\reff{eq:pdf tx rx 2}.

We have that $P[{\rm error}~2| {\rm encoding\,successful}]\to 0$ when $N \goes \infty$ if:
\begin{itemize}
\item
When the event $E_{2,1}$ occurs we have $\wh_{2c}\neq 1$.
In this case the received $Y_2^N$ is independent of the transmitted sequences.
This follows from the fact that the codewords $U_{2c}^N$ are generated in an
iid fashion and all the other codewords are generated independently
conditioned on $U_{2c}^N$. Hence, when decoder~2 finds a wrong $U_{2c}^N$,
all the decoded codewords are independent of the transmitted ones.
We can bound the error probability of $E_{2,1}$ as:
\begin{align*}
&P[E_{2,1}]
 = P \lsb \bigcup_{\wt_{2c}\neq 1 , \wt_{2pa}, \wt_{1c},\wt_{2pb},b_0, b_2} \rnone
\\
&\lnone (Y_2^N,U_{2c}^N(\wt_{2c}),U_{1c}^N(\wt_{1c}, \wt_{2c},b_0), X_{2}^N(\wt_{2c},\wt_{2pa}), U_{2pb}^N(\wt_{2c},\wt_{2pa},\wt_{1c},b_0,\wt_{2pb},b_2) )\in T_\epsilon^N \lb p_{Y_2, U_{2c},U_{1c},X_{2},U_{2pb}} \rb\rsb
\\
&\leq  2^{N(R_{2c}+R_{2pa}+R_{1c}+R'_{1c}  +R_{2pb}+R'_{2pb})}
\sum_{(y_2^N,u^N_{2c},u^N_{1c},x^N_{2},u^N_{2pb})\in T_\epsilon^N \lb p_{Y_2, U_{2c},U_{1c},X_{2},U_{2pb}} \rb}
p_{2|\star}|_{\star=\emptyset}
\\
&\leq  2^{N(R_{2c}+R_{2pa}+R_{1c}+R'_{1c}  +R_{2pb}+R'_{2pb}- I_{2|\star}|_{\star=\emptyset})}
\end{align*}
for $p_{2|\star}$ given in~\reff{eq:pdf dec 2} and $I_{2|\star}$ given in~\reff{eq:mutual info 2}.
Hence $P[E_{2,1}]\to 0$ as $N \goes \infty$ if~\reff{eq:our achievable region Ed2:1} is satisfied.

\item
When the event $E_{2,2}$ occurs, i.e., either $E_{2,2a}$ or $E_{2,2b}$, we have $\wh_{2c}=1$ but $\wh_{2pa}\neq 1$.
Whether $\wh_{1c}$ is correct or not, it does not matter since decoder~2 is not interested in $\wh_{1c}$.
However we need to consider whether the pair $(\wh_{1c},\bh_0)$ is equal to the transmitted one or not because this
affects the way the joint probability among all involved RVs factorizes.
We have:
\begin{itemize}
\item
Case $E_{2,2a}$: either $\wh_{1c} \neq 1$ or $\bh_0\neq \bo_0$.
In this case, conditioned on the (correct) decoded sequence $U_{2c}^N$,
the output $Y_2^N$ is independent of the (wrong) decoded sequences $U_{1c}^N$,
$X_2^N$ and also of $U_{2pb}^N$ (because $U_{2pb}^N$ is superimposed to the wrong pair $(U_{1c}^N,X_2^N)$).
It is easy to see that the most stringent error event is when both $\wh_{1c} \neq 1$ and $\bh_0\neq \bo_0$.
Thus we have
\begin{align*}
&P[E_{2,2a}]
 = P \lsb \bigcup_{\wt_{2pa}\neq 1, \wt_{1c}\neq 1,b_0\neq \bo_0, \wt_{2pb},b_2} \rnone
\\
&\lnone (Y_2^N,U_{2c}^N(1),U_{1c}^N(1, \wt_{1c},b_0), X_{2}^N(1, \wt_{2pa}), U_{2pb}^N(1, \wt_{2pa}, \wt_{1c},b_0, \wt_{2pb},b_2) )\in T_\epsilon^N \lb p_{Y_2, U_{2c},U_{1c},X_{2},U_{2pb}} \rb\rsb
\\
&\leq  2^{N(R_{2pa}+R_{1c}+R'_{1c}  +R_{2pb}+R'_{2pb})}
\sum_{(y_2^N,u^N_{2c},u^N_{1c},x^N_{2},u^N_{2pb})\in T_\epsilon^N \lb p_{Y_2, U_{2c},U_{1c},X_{2},U_{2pb}} \rb}
p_{2|\star}|_{\star=U_{2c}}
\\
&\leq  2^{N(R_{2pa}+R_{1c}+R'_{1c}  +R_{2pb}+R'_{2pb}- I_{2|\star}|_{\star=U_{2c}})}
\end{align*}
for $p_{2|\star}$ given in~\reff{eq:pdf dec 2} and $I_{2|\star}$ given in~\reff{eq:mutual info 2}.
Hence $P[E_{2,2a}] \to 0$ as $N \goes \infty$ if~\reff{eq:our achievable region Ed2:2a} is satisfied.

\item
Case $E_{2,2b}$: both $\wh_{1c} = 1$ and $\bh_0= \bo_0$.
In this case, conditioned on the (correct) decoded $(U_{2c}^N,U_{1c}^N)$,
the output $Y_2^N$ is independent of the (wrong) decoded sequences $(X_{2}^N,U_{2pb}^N)$. Thus we have
\begin{align*}
&P[E_{2,2b}]
 = P \lsb \bigcup_{\wt_{2pa}\neq 1, \wt_{2pb},b_2} \rnone
\\
&\lnone (Y_2^N,U_{2c}^N(1),U_{1c}^N(1,1,\bo_0), X_{2}^N(1, \wt_{2pa}), U_{2pb}^N(1, \wt_{2pa}, 1, \bo_0, \wt_{2pb},b_2) )\in T_\epsilon^N \lb p_{Y_2, U_{2c},U_{1c},X_{2},U_{2pb}} \rb\rsb
\\
&\leq  2^{N(R_{2pa}+R_{2pb}+R'_{2pb})}
\sum_{(y_2^N,u^N_{2c},u^N_{1c},x^N_{2},u^N_{2pb})\in T_\epsilon^N \lb p_{Y_2, U_{2c},U_{1c},X_{2},U_{2pb}} \rb}
p_{2|\star}|_{\star=(U_{2c},U_{1c})}
\\
&\leq  2^{N(R_{2pa}+R_{2pb}+R'_{2pb}- I_{2|\star}|_{\star=(U_{2c},U_{1c})})}
\end{align*}
for $p_{2|\star}$ given in~\reff{eq:pdf dec 2} and $I_{2|\star}$ given in~\reff{eq:mutual info 2}.
Hence $P[E_{2,2b}] \to 0$ as $N \goes \infty$ if~\reff{eq:our achievable region Ed2:2b} is satisfied.
\end{itemize}

\item
When the event $E_{2,3}$ occurs, i.e., either $E_{2,3a}$ or $E_{2,3b}$, we have $\wh_{2c}=1$,$\wh_{2pa}=1$ but $\wh_{2pb}\neq 1$.
Again, whether $\wh_{1c}$ is correct or not, it does not matter since decoder~2 is not interested in $\wh_{1c}$.
However we need to consider whether the pair $(\wh_{1c},\bh_0)$ is equal to the transmitted one or not because this
affects the way the joint probability among all involved RVs factorizes.
The analysis proceeds as for the event $E_{2,2}$.

We have:
\begin{itemize}
\item
Case $E_{2,3a}$: either $\wh_{1c} \neq 1$ or $\bh_0\neq \bo_0$.
In this case, conditioned on the (correct) decoded sequences $(U_{2c}^N,X_{2}^N)$,
the output $Y_2^N$ is independent of the (wrong) decoded sequences $(U_{1c}^N,U_{2c}^nU_{2pb}^N)$.
It is easy to see that the most stringent error event is when both $\wh_{1c} \neq 1$ and $\bh_0\neq \bo_0$.
Thus we have
\begin{align*}
&P[E_{2,3a}]
 = P \lsb \bigcup_{\wt_{1c}\neq 1,b_0\neq \bo_0, \wt_{2pb},b_2} \rnone
\\
&\lnone (Y_2^N,U_{2c}^N(1),U_{1c}^N(1, \wt_{1c},b_0), X_{2}^N(1, 1), U_{2pb}^N(1, 1, \wt_{1c},b_0, \wt_{2pb},b_2) )\in T_\epsilon^N \lb p_{Y_2, U_{2c},U_{1c},X_{2},U_{2pb}} \rb\rsb
\\
&\leq  2^{N(R_{1c}+R'_{1c}  +R_{2pb}+R'_{2pb})}
\sum_{(y_2^N,u^N_{2c},u^N_{1c},x^N_{2},u^N_{2pb})\in T_\epsilon^N \lb p_{Y_2, U_{2c},U_{1c},X_{2},U_{2pb}} \rb}
p_{2|\star}|_{\star=(U_{2c},X_{2})}
\\
&\leq  2^{N(R_{2pa}+R_{1c}+R'_{1c}  +R_{2pb}+R'_{2pb}- I_{2|\star}|_{\star=(U_{2c},X_{2})})}
\end{align*}
for $p_{2|\star}$ given in~\reff{eq:pdf dec 2} and $I_{2|\star}$ given in~\reff{eq:mutual info 2}.
Hence $P[E_{2,3a}] \to 0$ as $N \goes \infty$ if~\reff{eq:our achievable region Ed2:3a} is satisfied.

\item
Case $E_{2,3b}$: both $\wh_{1c} = 1$ and $\bh_0= \bo_0$.
In this case, conditioned on the (correct) decoded sequences $(U_{2c}^N,X_{2}^N,U_{1c}^N)$,
the output $Y_2^N$ is independent of the (wrong) decoded sequence $U_{2pb}^N$.
However, since $(U_{2c}^N,X_{2}^N,U_{1c}^N)$ is the triplet that passed the encoding binning step,
they are jointly typical. Hence, in this case we cannot use the factorization in $p_{2|\star}$
given in~\reff{eq:pdf dec 2}, but we need to replace $p_{U_{1c}|U_{2c}}$ in~\reff{eq:pdf dec 2}
with $p_{U_{1c}|U_{2c},X_{2}}$. Thus we have
\begin{align*}
&P[E_{2,3b}]
 = P \lsb \bigcup_{\wt_{2pb},b_2} \rnone
\\
&\lnone (Y_2^N,U_{2c}^N(1),U_{1c}^N(1,1,\bo_0), X_{2}^N(1, 1), U_{2pb}^N(1, 1, 1, \bo_0, \wt_{2pb},b_2) )\in T_\epsilon^N \lb p_{Y_2, U_{2c},U_{1c},X_{2},U_{2pb}} \rb\rsb
\\
&\leq  2^{N(R_{2pb}+R'_{2pb})}
\sum_{(y_2^N,u^N_{2c},u^N_{1c},x^N_{2},u^N_{2pb})\in T_\epsilon^N \lb p_{Y_2, U_{2c},U_{1c},X_{2},U_{2pb}} \rb}
p_{U_{2c}}\,p_{X_{2}|U_{2c}}\, p_{U_{1c}|U_{2c},X_2}\, p_{U_{2pb}|U_{2c},U_{1c},X_{2}}\,p_{Y_2|U_{1c},U_{2c},X_2}
\\
&\leq  2^{N(R_{2pb}+R'_{2pb}-  I(Y_2; U_{2pb}|U_{1c},U_{2c},X_2))}
\end{align*}
Hence $P[E_{2,3b}] \to 0$ as $N \goes \infty$ if~\reff{eq:our achievable region Ed2:3b} is satisfied.

\end{itemize}

\end{itemize}

\subsection{Decoding Errors at Decoder 1}

\begin{table}
  \centering
  \caption{Error events at decoder 1.}\label{tab:Errors1}
\begin{tabular}{|l|l|l|l|l|}
\hline
Event     & $w_{2c}$ & $(w_{1c},b_1)$ & $w_{1pb}$ & $p_{Y_1|\star}$ \\\hline
$E_{1,1}$ & X &$\cdots$&$\cdots$& $p_{Y_1}$ \\ \hline
$E_{1,2}$ & 1 & X &$\cdots$& $p_{Y_1|U_{2c}}$ \\ \hline
$E_{1,3}$ & 1 & 1 & X & $P_{Y_1|U_{2c},U_{1c}}$ \\\hline
\end{tabular}
\end{table}

The probability of error at decoder~1 is bounded as:
\[
P[{\rm error}~1| {\rm encoding\,successful}]\leq \sum_{i=1}^3 P[E_{1,i}],
\]
where $P[E_{1,i}]$ is the error event defined in Table~\ref{tab:Errors1}.
The meaning of the symbols in Table~\ref{tab:Errors1} is as for Table~\ref{tab:Errors2}.
We have that $P[{\rm error}~1| {\rm encoding\,successful}]\to 0$ when $N \goes \infty$ if:
\begin{itemize}
\item  When the event $E_{1,1}$ occurs we have $\wh_{2c}\neq 1$.
In this case the received $Y_1^N$ is independent of the transmitted sequences.
We can bound the error probability of $E_{1,1}$ as:
\begin{align*}
&P[E_{1,1}]
 = P \lsb \bigcup_{\wt_{2c}\neq 1 , \wt_{1c},\wt_{1pb},b_0, b_1} \rnone
\\
&\lnone (Y_1^N,U_{2c}^N(\wt_{2c}),U_{1c}^N(\wt_{1c}, \wt_{2c},b_0), U_{1pb}^N(\wt_{2c},\wt_{2pa},\wt_{1c},b_0,\wt_{2pb},b_1) )\in T_\epsilon^N \lb p_{Y_1, U_{2c},U_{1c},U_{1pb}} \rb\rsb
\\
&\leq  2^{N(R_{2c}+R_{1c}+R'_{1c}  +R_{1pb}+R'_{1pb} )}
\sum_{(y_1^N,u^N_{2c},u^N_{1c},u^N_{1pb})\in T_\epsilon^N \lb p_{Y_1, U_{2c},U_{1c},U_{1pb}} \rb}
p_{1|\star}|_{\star=\emptyset}
\\
&\leq  2^{N(R_{2c}+R_{2pa}+R_{1c}+R'_{1c}  +R_{2pb}+R'_{2pb}- I_{1|\star}|_{\star=\emptyset})}
\end{align*}
for $p_{1|\star}$ given in~\reff{eq:pdf dec 2} and $I_{1|\star}$ given in~\reff{eq:mutual info 1}.
Hence $P[E_{1,1}]\to 0$ as $N \goes \infty$ if~\reff{eq:our achievable region Ed1:1} is satisfied.

\item
When the event $E_{1,2}$ occurs,  either $\wh_{1c} \neq 1$, $\bh_0\neq \bo_0$ or both.
In this case, conditioned on the (correct) decoded sequence $U_{2c}^N$,
the output $Y_1^N$ is independent of the (wrong) decoded sequences $U_{1c}^N$ and $U_{1pb}^N$ .
It is easy to see that the most stringent error event is when both $\wh_{1c} \neq 1$ and $\bh_0\neq \bo_0$.
Thus we have
\begin{align*}
&P[E_{1,2}]
 = P \lsb \bigcup_{\wt_{1c}\neq 1,b_0\neq \bo_0, \wt_{1pb},b_1} \rnone
\\
&\lnone (Y_1^N,U_{2c}^N(1),U_{1c}^N(1, \wt_{1c},b_0), U_{1pb}^N(1, \wt_{1c},b_0, \wt_{1pb},b_1) )\in T_\epsilon^N \lb p_{Y_1, U_{2c},U_{1c},U_{1pb}} \rb\rsb
\\
&\leq  2^{N(R_{1c}+R'_{1c}  +R_{1pb}+R'_{1pb} )}
\sum_{(y_1^N,u^N_{2c},u^N_{1c},u^N_{1pb})\in T_\epsilon^N \lb p_{Y_1, U_{2c},U_{1c},U_{1pb}} \rb}
p_{1|\star}|_{\star=U_{2c}}
\\
&\leq  2^{N(R_{1c}+R'_{1c}  +R_{1pb}+R'_{1pb} - I_{1|\star}|_{\star=U_{2c}})}
\end{align*}
for $p_{1|\star}$ given in~\reff{eq:pdf dec 1} and $I_{1|\star}$ given in~\reff{eq:mutual info 1}.
Hence $P[E_{1,2}] \to 0$ as $N \goes \infty$ if~\reff{eq:our achievable region Ed1:2} is satisfied.

\item

When the event $E_{1,3}$ occurs,  either $\wh_{1pb} \neq 1$, $\bh_1\neq \bo_1$ or both.
In this case, conditioned on the (correct) decoded sequence $U_{2c}^N$ and $U_{1c}^N)$,
the output $Y_1^N$ is independent of the (wrong) decoded sequences  $U_{1pb}^N$.
It is easy to see that the most stringent error event is when both $\wh_{1pb} \neq 1$ and $\bh_1\neq \bo_1$.
Thus we have
\begin{align*}
&P[E_{1,3}]
 = P \lsb \bigcup_{\wt_{1pb}\neq 1,b_1\neq \bo_1} \rnone
\\
&\lnone (Y_1^N,U_{2c}^N(1),U_{1c}^N(1, 1,\bo_0), U_{1pb}^N(1, 1,\bo_0, \wt_{1pb},b_1) )\in T_\epsilon^N \lb p_{Y_1, U_{2c},U_{1c},U_{1pb}} \rb\rsb
\\
&\leq  2^{N(R_{1pb}+R'_{1pb} )}
\sum_{(y_1^N,u^N_{2c},u^N_{1c},u^N_{1pb})\in T_\epsilon^N \lb p_{Y_1, U_{2c},U_{1c},U_{1pb}} \rb}
p_{1|\star}|_{\star=U_{2c},U_{1c}}
\\
&\leq  2^{N(R_{1c}+R'_{1c}  +R_{1pb}+R'_{1pb} - I_{1|\star}|_{\star=U_{2c},U_{1c}})}
\end{align*}
for $p_{1|\star}$ given in~\reff{eq:pdf dec 1} and $I_{1|\star}$ given in~\reff{eq:mutual info 1}.
Hence $P[E_{1,3}] \to 0$ as $N \goes \infty$ if~\reff{eq:our achievable region Ed1:3} is satisfied.

\end{itemize}



\subsection{Proof of Lemma \ref{thm:two step binnig} }
\label{app:proof two step binnig}
An encoding error is committed if we cannot find a $b_0$ in the first step or if, upon finding  the correct $b_0$ in the first encoding step, we cannot find the correct $(b_1,b_2)$ in  the second step.  Let $E_{e,0}$  the probability of the first event and $E_{e,12}$ of the latter, than:

\pp{
P[{\rm encoding\,NOT\,successful}] \leq P[E_{e,0}]+P[E_{e,12}| E_{e,0}^c]
}
where
\pp{
P[E_{e,0}] &= P[\bigcap_{b_0=1}^{2^{N R'_{1c}   }}  \lb U_{2c}^N(1), X_{2}^N(1,1),U_{1c}^N(1,1,b_0) \rb  \notin T_\epsilon^N (p_{U_{2c},X_{2},U_{1c}}) ] \\
&=(1-P[\lb U_{2c}^N(1), X_{2}^N(1,1),U_{1c}^N(1,1,b_0) \rb  \notin T_\epsilon^N (p_{U_{2c},X_{2},U_{1c}}) ])^{2^{N R'_{1c}  }}.
}
Using standard  typicality arguments we have
\pp{
P[\lb U_{2c}^N(1), X_{2}^N(1,1),U_{1c}^N(1,1,b_0) \rb  \notin T_\epsilon^N (p_{U_{2c},X_{2},U_{1c}}) ] & =\sum_{u_{1c} \in T_\epsilon^N (p_{U_{2c},X_{2},U_{1c}}|U_{2c},X_2)}
&\geq (1-\ep) 2^{N(I(U_{1c}; X_2|U_{2c})+\de)}.
}

Now we can write
\pp{
P[E_{e,0}] &\leq (1-(1-\ep) 2^{N(I(U_{1c}; X_2|U_{2c})+\de)}) ^{2^{N R'_{1c}  }} \\
&\leq \exp \lb 1-(1-\ep) 2^{N(R'_{1c}  -I(U_{1c}; X_2|U_{2c})+\de)}) \rb
}
so that  $P[E_{e,0}] \goes 1$ when $N \goes 0$ if \reff{eq: two step binning R"1c} is satisfied.

Now the error event $E_{e,12}$ can be divided in three distinct error events:
\begin{itemize}
  \item $E_{e,21 \ a}$: it is not possible to find $b_1$ such that
  $
  (U_{2c}^N,X_{2}^N, U_{1c}^N,U_{1pb}^N) \in T_\epsilon^N(  p_{U_{2c},X_{2},U_{1c},U_{1pb}}),
  $
  \item $E_{e,21 \ b}$: it is not possible to find $b_2$ such that
  $
  (U_{2c}^N,X_{2}^N, U_{1c}^N,U_{2pb}^N)
 \in T_\epsilon^N(  p_{U_{2c},X_{2},U_{1c},U_{2pb}}).
  $
  \item $E_{e,21 \ c}$ Given that we can find $b_1$ and $b_2$  satisfy the first two equations, we cannot find a couple $(b_1,b_2)$ such that
  $
  (U_{2c}^N,X_{2}^N, U_{1c}^N,U_{1pb}^N,U_{2pb}^N)
 \in T_\epsilon^N(  p_{U_{2c},X_{2},U_{1c},U_{1pb},U_{2pb}}).
  $
\end{itemize}
We now  establish the rate bounds that guarantee that the probability of error of each of these events goes to zero.

For $E_{e,21 \ a}$ we have:
\pp{
P [E_{e,21 \ a}] &= (1-P[\lb U_{2c}^N(1), X_{2}^N(1,1),U_{1c}^N(1,1,b_0),U_{1pb}^N(1,1,b_0,1,b_1) \rb  \notin T_\epsilon^N (p_{U_{2c},X_{2},U_{1c},U_{1pb}})])^{2^{N R'_{1pb} }},
}
where
$$
P[\lb U_{2c}^N(1), X_{2}^N(1,1),U_{1c}^N(1,1,b_0),U_{1pb}^N(1,1,b_0,1,b_1) \rb
\notin T_\epsilon^N (p_{U_{2c},X_{2},U_{1c},U_{1pb}})] \geq (1-\ep) 2^{-N(I(X_2; U_{1pb}| U_{2c},U_{1c})+\de)}.
$$
As for $E_{e,0}$, this implies that $P[E_{e,21 \ a}] \goes 1$ when $N \goes 0$ if \reff{eq: two step binning R"1pb} is satisfied.

For $E_{e,21 \ b}$, we have that the probability of this event goes to one for large $N$ given that $(U_{2c},X_2,U_{1c})$ appear to be generated according to the distribution
$p_{U_{2c},X_2,U_{1c}}$ and $U_{2pb}$ is generated according to $p_{U_{2pb}| U_{2c}, X_2, U_{1c}}$.

For $E_{e,21 \ c}$ we have:
\pp{
P [E_{e,21 \ c}] &= (1-P[\lb U_{2c}^N(1), X_{2}^N(1,1),U_{1c}^N(1,1,b_0),U_{1pb}^N(1,1,b_0,1,b_1),U_{1pb}^N(1,1,b_0,1,b_2)  \rb \\
& \lag \notin T_\epsilon^N (p_{U_{2c},X_{2},U_{1c},U_{1pb},U_{2pb}})])^{2^{N (R'_{1pb} +R'_{2pb})}},
}
where
$$
P[\lb U_{2c}^N(1), X_{2}^N(1,1),U_{1c}^N(1,1,b_0),U_{1pb}^N(1,1,b_0,1,b_1),U_{1pb}^N(1,1,b_0,1,b_2)  \rb \\
\notin T_\epsilon^N (p_{U_{2c},X_{2},U_{1c},U_{1pb},U_{2pb}})]) \leq 2^{I()+\de}.
$$
This implies that $P[E_{e,21 \ c}] \goes 1$ when $N \goes 0$ if \reff{eq: two step binning R1pb+R"2pb} is satisfied.

\subsection{Containment of \cite[Thm. 1]{DevroyeThesis} in $\Rcal_{RTD}$ of Section \ref{sec:devroye}}
\label{sec:devroye thesis achievable region}
We refer to the region in \cite[Thm. 1]{DevroyeThesis} as $\Rcal_{DMT}$ for brevity. We show this inclusion of $\Rcal_{DMT}$ in $\Rcal_{RTD}$ with the following steps:\\
\noindent
$\bullet$ We enlarge the region ${\cal R}_{DMT}$ by removing some rate constraints. \\
$\bullet$ We further enlarge the region by enlarging the set of possible input distributions. This allows us to remove the $V_{11}$  and $Q$ from the inner bound. We refer to this region as $\Rcal_{DMT}^{out}$  since is enlarges the original achievable region.\\
$\bullet$ We make a correspondence between the RVs and corresponding rates of $\Rcal_{DMT}^{out}$ and ${\cal R}_{RTD}$.  \\
$\bullet$ We choose a particular subset of ${\cal R}_{RTD}$, $\Rcal_{RTD}^{in}$, for which we can more easily show  $\Rcal_{DMT}\subseteq \Rcal_{DMT}^{out} \subset \Rcal_{RTD}^{in} \subseteq \Rcal_{RTD}$, since
 $\Rcal_{DMT}^{out}$ and $\Rcal_{RTD}^{in}$  have  identical input distribution decompositions and similar rate bound equations.

 {\bf Enlarge the region ${\cal R}_{DMT}$}\\
We first enlarge the rate region of \cite[Thm. 1]{DevroyeThesis}, $\Rcal_{DMT}$ by removing a number of constraints
(specifically, we remove equations (2.6, 2.8, 2.10, 2.13, 2.14, 2.16 2.17) of \cite[Thm. 1]{DevroyeThesis}).
%
Also, following the line of thoughts in  \cite[Appendix D]{willems1985discrete} it is possible to show that without loss of generality we can set $X_1$
 to be a deterministic function of $V_{11}$ and $V_{12}$, allowing us insert $X_1$ next to $V_{11},V_{12}$. With these consideration we can enlarge the original region
 and define $ \Rcal_{DMT}^{out} $ as follows.
\begin{subequations}
\ea{
R_{21}' &=& I(V_{21}; X_1, V_{11}, V_{12}|W)
\label{eq: devroye 1}\\
R_{22}' &=& I(V_{22}; X_1, V_{11}, V_{12}|W)
\label{eq: devroye 2}\\
R_{11} &\leq&   I(Y_1,V_{12},V_{21}; V_{11}|W)
\label{eq: devroye 3}\\
R_{21}+R_{21}' &\leq& I(Y_1,X_1,V_{11},V_{12};V_{21}|W)
\label{eq: devroye 4}\\
R_{11}+R_{21}+R_{21}' &\leq& I(Y_1,V_{12};V_{11},V_{21}|W)+I(V_{11};V_{21}|W)
\label{eq: devroye 5}\\
R_{11}+R_{21}+R_{21}'+R_{12} &\leq& I(Y_1;X_1,V_{11},V_{12},V_{21}|W)+I(X_1,V_{11},V_{12};V_{21}|W)
\label{eq: devroye 6}\\
R_{22}+R_{22}' &\leq& I(Y_2,V_{12},V_{21}; V_{22}|W)
\label{eq: devroye 7}\\
R_{22}+R_{22}'+R_{21}+R_{21}' &\leq& I(Y_2,V_{12}; V_{22},V_{21}|W)+I(V_{22};V_{21}|W)
\label{eq: devroye 8}\\
R_{22}+R_{22}'+R_{21}+R_{21}' +R_{12} &\leq& I(Y_2; V_{22},V_{21},V_{12}|W)+I(V_{22},V_{21};V_{12}|W)
\label{eq: devroye 9}
}
\end{subequations}
taken over the union of  distributions
\ea{
p_{W}p_{V_{11}}p_{V_{12}} p_{X_1|V_{11},V_{12}} p_{V_{21}| V_{11} V_{12}} p_{V_{22}| V_{11},V_{12}} p_{X_2| V_{11},V_{12}, V_{21}, V_{22}}.
\label{eq: devroye factorization}
}

Using the factorization of the auxiliary RVs in \cite[Thm. 1]{DevroyeThesis},  we may insert $X_1$ next to $V_{11}$  in equation \eqref{eq: devroye 6}. 

For equation (\ref{eq: devroye 3}):
\pp{
R_{11} &\leq&   I(Y_1,V_{12},V_{21}; V_{11}|W)\\
& = & I(Y_1,V_{21}; V_{11}|V_{12},W)+I(V_{12};V_{11}|W)\\
& = & I(Y_1,V_{21}; V_{11}|V_{12},W)\\
& = & I(Y_1,V_{21}; X_1,V_{11}|V_{12},W)\\
& = & I(Y_1;X_1, V_{11}| V_{12},V_{21},W)+I(V_{21}; X_1, V_{11} |V_{12},W).\\
}
For equation (\ref{eq: devroye 5}) we have:
\pp{
R_{11}+R_{21}+R_{21}' &\leq& I(Y_1,V_{12};V_{11},V_{21}|W)+I(V_{11};V_{21}|W)\\
& = & I(Y_1;V_{11},V_{21}|V_{12},W)+I(V_{12};V_{11},V_{21}|W)+I(V_{11};V_{21}|W)\\
& = & I(Y_1;V_{11},V_{21}|V_{12},W)+I(V_{12};V_{21}|V_{11},W)+I(V_{11};V_{21}|W)\\
& = & I(Y_1;V_{11},V_{21}|V_{12},W)+I(V_{11},V_{12};V_{21}|W)\\
& = & I(Y_1;X_1,V_{11},V_{21}|V_{12},W)+I(X_1,V_{11},V_{12};V_{21}|W)\\
}

The original region is thus equivalent to
\begin{subequations}
\ea{
R_{21}' &=& I(V_{21}; X_1,V_{11}, V_{12}|W)\\
R_{22}' &=& I(V_{22}; X_1,V_{11}, V_{12}|W)\\
R_{11} &\leq&  I(Y_1;X_1,V_{11}| V_{12},V_{21}|W)+I(V_{21}; X_1 |V_{12},W)\\
R_{21}+R_{21}' &\leq& I(Y_1,X_1,V_{11},V_{12};V_{21}|W)\\
R_{11}+R_{21}+R_{21}' &\leq&I(Y_1;X_1,V_{11},V_{21}|V_{12},W)+I(X_1;V_{21}|W)\\
R_{11}+R_{21}+R_{21}'+R_{12} &\leq& I(Y_1;X_1,V_{11},V_{21},V_{12}|W)+I(X_1,V_{11},V_{12};V_{21}|W)\\
R_{22}+R_{22}' &\leq& I(Y_2,V_{12},V_{21}; V_{22}|W)\\
R_{22}+R_{22}'+R_{21}+R_{21}' &\leq& I(Y_2,V_{12}; V_{22},V_{21}|W)+I(V_{22};V_{21}|W)\\
R_{22}+R_{22}'+R_{21}+R_{21}' +R_{12} &\leq& I(Y_2; V_{22},V_{21},V_{12}|W)+I(V_{22},V_{21};V_{12}|W)
}
\label{eq:devroye step 1}
\end{subequations}
union over all  distributions that factor as in \reff{eq: devroye factorization}.

\noindent
{\bf Enlarge the  class of input distribution and eliminate $V_{11}$ and $W$ }\\
Now increase the set of possible input distributions of equation \ref{eq: devroye factorization} by letting $V_{11}$ have any joint distribution with $V_{12}$. This is done by substituting $p_{V_{11}}$ with $p_{V_{11}| V_{12}}$ in the expression of the input distribution.  With this substitution we have:
\pp{
p_{W}p_{V_{11}|V_{12}}p_{V_{12}} p_{X_1|V_{11},V_{12}} p_{V_{21}| X_{1},V_{11} V_{12}} p_{V_{22}| X_{1},V_{11},V_{12}} p_{X_2| X_1,V_{11},V_{12}, V_{21}, V_{22}}\\
= \; p_{W}p_{V_{12}} p_{V_{11},X_1|V_{12}} p_{V_{21}| X_{1},V_{11} V_{12}} p_{V_{22}| X_{1},V_{11},V_{12}} p_{X_2| X_1,V_{11},V_{12}, V_{21}, V_{22}}\\
= \;  p_{W}p_{V_{12}} p_{X_1'|V_{12}} p_{V_{21}| X_1', V_{12}} p_{V_{22}| X_1',V_{12}} p_{X_2| X_1',V_{12}, V_{21}, V_{22}}\\
}
with $X_1'=(X_1,V_{11})$.   Since $V_{12}$ is decoded at both decoders, the time sharing random $W$ may be
incorporated with $V_{12}$ without loss of generality and thus can be dropped.
The region described in (\ref{eq:devroye step 1}) is convex and thus time sharing is not needed.
With these simplifications, the region $\Rcal_{DMT}^{out}$ is now defined as
\begin{subequations}
\ea{
R_{21}' &=& I(V_{21}; X_1', V_{12})\\
R_{22}' &=& I(V_{22}; X_1', V_{12})\\
R_{11} &\leq&  I(Y_1;X_1'| V_{12},V_{21})+I(V_{21}; X_1 |V_{12})\\
R_{21}+R_{21}' &\leq& I(Y_1,X_1',V_{12};V_{21})\\
R_{11}+R_{21}+R_{21}' &\leq&I(Y_1;X_1',V_{21}|V_{12})+I(X_1;V_{21})\\
R_{11}+R_{21}+R_{21}'+R_{12} &\leq& I(Y_1;X_1',V_{21},V_{12})+I(X_1',V_{12};V_{21})\\
R_{22}+R_{22}' &\leq& I(Y_2,V_{12},V_{21}; V_{22})\\
R_{22}+R_{22}'+R_{21}+R_{21}' &\leq& I(Y_2,V_{12}; V_{22},V_{21})+I(V_{22};V_{21})\\
R_{22}+R_{22}'+R_{21}+R_{21}' +R_{12} &\leq& I(Y_2; V_{22},V_{21},V_{12})+I(V_{22},V_{21};V_{12})
}
\end{subequations}taken over the union of all distributions
$$
p_{V_{12}} p_{X_1'|V_{12}} p_{V_{21}| X_1', V_{12}} p_{V_{22}| X_1',V_{12}} p_{X_2| X_1',V_{12}, V_{21}, V_{22}}.
$$

 {\bf Correspondence between the random variables and rates.}
When referring to \cite{DevroyeThesis} please note that the index of the primary and cognitive user are reversed with respect to our notation (i.e $1 \goes 2$ and vice-versa). Consider the correspondences between the variables of \cite[Thm. 1]{DevroyeThesis} and those of Theorem \ref{thm:our achievable region} in Table \ref{tab:sec:devroye thesis achievable region} to obtain the region ${\cal R}_{DMT}^{out}$ defined as the set of rate pairs satisfying
\begin{table}
\centering
\caption{Assignment of RVs of Appendix  \ref{sec:devroye thesis achievable region} }\label{tab:sec:devroye thesis achievable region}
\begin{tabular}{| lll |} \hline
 RV, rate of Theorem \ref{thm:our achievable region}  &  RV, rate of  \cite[Thm. 1]{DevroyeThesis} & Comments \\ \hline
$U_{2c}, R_{2c}$ &$ V_{12}, R_{12}$ & TX 2 $\goes$ RX 1, RX 2\\  
$U_{1c}, R_{1c}$ & $V_{21}, R_{21}$ & TX 1 $\goes$ RX 1, RX 2\\  
$U_{1pb}, R_{1pb}$ & $V_{22}, R_{22}$ & TX 1 $\goes$ RX 1\\  
$X_2, R_{2pa}$ & $X_1', R_{11}$ & TX 2 $\goes$ RX 2 \\
$U_{2pb}=\emptyset, R_{2pb}'=0$ & -- &  TX 1 $\goes$  RX 2\\  
$R_{1c}' = I(U_{1c};X_2|U_{2c})$ & $L_{21}-R_{21} = I(V_{21};V_{11},V_{12})$ & Binning rate  \\ 
$R_{1pb}' = I(U_{1pb};X_2|U_{1c},U_{2c} )$ & $L_{22}-R_{22}  = I(V_{22};V_{11},V_{12})$ & Binning rate \\ 
$X_1$                 & $X_2$               &  \\
\hline
\end{tabular}
\end{table}

\begin{subequations}
\ea{
R_{1c}' &=& I(U_{1c}; X_2, U_{2c})
\label{e20}\\
R_{1pb}' &=& I(U_{1pb}; X_2, U_{2c})
\label{e21}\\
R_{2pa}+R_{1c}+R_{1c}'+R_{2c} &\leq& I(Y_2;U_{1c},U_{2c},X_2)+I(X_2,U_{2c};U_{1c})
\label{e23}\\
R_{2pa}+R_{1c}+R_{1c}' &\leq&I(Y_2;X_2,U_{1c}|U_{2c})+I(X_2;U_{1c})
\label{e24}\\
R_{1c}+R_{1c}' &\leq& I(Y_2,X_2,U_{2c};U_{1c})
\label{e25}\\
R_{2pa} &\leq&  I(Y_2;X_2| U_{2c},U_{1c})+I(U_{1c}; X_2 |U_{2c})
\label{e26}\\
R_{1pb}+R_{1pb}'+R_{1c}+R_{1c}' +R_{2c} &\leq& I(Y_1; U_{1pb},U_{1c},U_{2c})+I(U_{1pb},U_{1c};U_{2c})
\label{e27}\\
R_{1c}+R_{1pb}+R_{1c}'+R_{1pb}' &\leq& I(Y_1,U_{2c}; U_{1pb},U_{1c})+I(U_{1pb};U_{1c})
\label{e28}\\
R_{1pb}+R_{1pb}' &\leq& I(Y_1,U_{2c},U_{1c}; U_{1pb})
\label{e29}
}
\end{subequations}
taken over the union of all distributions
\ea{
p_{U_{2c}} p_{X_2|U_{2c}} p_{U_{1c}| X_2} p_{U_{1pb}| X_2} p_{X_1| X_2, U_{1c}, U_{1pb}}.
\label{eq:factorDevroye}
}

Next, we using the correspondences of the table and restrict the fully general input distribution of Theorem
 \ref{thm:our achievable region} to match the more constrained factorization of (\ref{eq:factorDevroye}), obtaining a region  $ \Rcal_{RTD}^{in} \subseteq \Rcal_{RTD}$ defined as the set of rate tuples satisfying
\begin{subequations}
\ea{
R_{1c}'&=&I(U_{1c}; X_2|U_{2c})
\label{e10}\\                                                     
R_{1c}'+R_{1pb}'&=& I(X_2; U_{1c}, U_{1pb}| U_{2c})
\label{e12}\\
R_{2c}+R_{1c}+R_{2pa}+R_{1c}' &\leq& I(Y_2; U_{2c},U_{1c},X_2)+I(U_{1c}; X_2 |U_{2c})
\label{e13}\\
R_{2pa}+R_{1c}+R_{1c}'&\leq& I(Y_2; U_{1c}, X_2| U_{2c})+I( U_{1c}; X_2| U_{2c})
\label{e14}\\
R_{1c}+R_{1c}' &\leq& I(Y_2; U_{1c}| U_{2c}, X_2)+I(U_{1c}; X_2| U_{2c})
\label{e15}\\
R_{2pa} &\leq&I(Y_2; X_2| U_{2c},U_{1c})+I(U_{1c}; X_2| U_{2c})
\label{e16}\\
R_{1pb}+R_{1pb}'+R_{1c}+R_{1c}' +R_{2c}  &\leq& I(Y_1; U_{2c},U_{1c},U_{1pb})
\label{e17}\\
R_{1c}+R_{1pb}+R_{1c}'+R_{1pb}' &\leq& I(Y_1; U_{1c},U_{1pb}|U_{2c})
\label{e18}\\
R_{1pb}+R_{1pb}' &\leq& I(Y_1; U_{1pb}|U_{2c},U_{1c})
\label{e19}
}
\end{subequations}
 union of all distributions that factor as
$$
p_{U_{2c},X_2}  p_{U_{1c}| X_2 }p_{U_{1pb}| X_2} p_{X_1 | X_2, U_{1c}, U_{1pb}}.
$$


\noindent
{\bf Equation-by-equation comparison.} We now show that
$ \Rcal_{DMT}^{out}  \subseteq \Rcal_{RTD}^{in}$
by fixing an input distribution (which are the same for these two regions) and comparing the rate regions equation by equation. We refer to the equation numbers directly, and look at the difference between the corresponding equations in the two new regions.

\begin{itemize}
\item  \eqref{e13}-\eqref{e10} vs \eqref{e23}-\eqref{e20}: Noting the cancelation / interplay between the binning rates, we see that
$$
\lb \eqref{e13}-\eqref{e10}  \rb - \lb  \eqref{e24}-\eqref{e20} \rb=0.
$$
\item \eqref{e14}-\eqref{e10} vs. \eqref{e24}-\eqref{e20}: \pp{
\lb \eqref{e14}-\eqref{e10} \rb - \lb \eqref{e24}-\eqref{e20} \rb\\
\lag  \lag =-I(X_2;U_{1c})+I(U_{1c}; X_2,U_{2c}) \\
\lag  \lag =I(U_{2c}; U_{1c}|X_2)\\
\lag  \lag =0
}
\item \eqref{e15}-\eqref{e10} vs. \eqref{e25}-\eqref{e20}: again noting the cancelations, 
\pp{
\lb \eqref{e15}-\eqref{e10}\rb - \lb \eqref{e25}-\eqref{e20} \rb=0
}
\item \eqref{e16} vs. \eqref{e26}: 
\pp{
\eqref{e16} -\eqref{e26}=0 \\
}
\item  \eqref{e17}-\eqref{e12} vs. 
\eqref{e27}-\eqref{e21}-\eqref{e20}
\pp{
(\eqref{e17}-\eqref{e12})- (\eqref{e27}-\eqref{e21}-\eqref{e20})\\
\lag =-I(X_2; U_{1c}, U_{1pb}| U_{2c})\\
\lag  \lag   -I(U_{1pb},U_{1c}; U_{2c})+I(U_{1c}; U_{2c}, X_2)+I(U_{1pb}; U_{2c}, X_2)\\
\lag =-I(U_{1pb}, U_{1c};X_2, U_{2c}) +I(U_{1c}; U_{2c}, X_2)+I(U_{1pb}; U_{2c}, X_2)\\
\lag= -I(U_{1pb};X_2, U_{2c}) -I(U_{1c}; X_2, U_{2c}| U_{1pb})+I(U_{1c}; U_{2c}, X_2)+I(U_{1pb}; U_{2c}, X_2)\\
\lag= -I(U_{1c}; X_2, U_{2c}| U_{1pb})+I(U_{1c}; U_{2c}, X_2)\\
\lag=- H(U_{1c}| U_{1pb})+ H(U_{1c}| X_2, U_{2c},U_{1pb})+H(U_{1c})-H(U_{1c}| X_2, U_{2c})\\
\lag=I(U_{1c}; U_{1pb}) >0
}
where we have used the fact that $U_{1c}$ and $U_{1pb}$ are conditionally independent given $(U_{2c}, X_2)$.
\item  $\eqref{e18}-\eqref{e12}$ vs. $\eqref{e28}-\eqref{e21}-\eqref{e20}$:
\pp{
(\eqref{e18}-\eqref{e12})-(\eqref{e28}-\eqref{e21}-\eqref{e20})  \\
\lag = -I(X_2; U_{1c}, U_{1pb}|U_{2c})-I(U_{2c}; U_{1c}, U_{1pb})+I(U_{1pb};U_{2c}, X_2) - I(U_{1pb};U_{1c})+I(U_{1c}; X_2,U_{2c}) \\
\lag = -I(X_2,U_{2c}; U_{1c}, U_{1pb})+I(U_{1pb};U_{2c}, X_2) - I(U_{1pb};U_{1c})+I(U_{1c}; X_2,U_{2c}) \\
\lag = -I(X_2,U_{2c}; U_{1pb})-I(U_{1c}; X_2,U_{2c}| U_{1pb})+I(U_{1pb};U_{2c}, X_2) - I(U_{1pb};U_{1c})+I(U_{1c}; X_2,U_{2c}) \\
\lag = -I(U_{1c}; X_2,U_{2c},U_{1pb}) +I(U_{1c}; X_2,U_{2c}) \\
\lag = -I(U_{1c}; X_2,U_{2c})-I(U_{1c}; U_{1pb}|X_2,U_{2c}) +I(U_{1c}; X_2,U_{2c}) \\
\lag = 0
}
where we have used the fact that $U_{1c}$ and $U_{1pb}$ are conditionally independent given $(U_{2c}, X_2)$.
\item  $\eqref{e19}-\eqref{e12} +\eqref{e10}$  vs. $\eqref{e29}-\eqref{e21}$:
\pp{
(\eqref{e19}-\eqref{e12} +\eqref{e10}) - (\eqref{e29}-\eqref{e21}) \\
\lag = -I(U_{1pb}; X_2| U_{2c}, U_{1c})-I(U_{1pb}; U_{2c}, U_{1c})+I(U_{1pb}; X_2, U_{2c}) \\
\lag = -I(U_{1pb}; X_2, U_{2c}, U_{1c} )+I(U_{1pb}; U_{2c}, X_2 )\\
\lag =- I(U_{1pb}; U_{1c} | U_{2c}, X_2) \\
\lag= 0
}
\end{itemize}

\bigskip

\subsection{Containment of \cite[Thm. 2]{biao2009Arxiv} in $\Rcal_{RTD}$ of Section \ref{sec:biao}}
\label{sec:biao}

The independently derived region in \cite[Thm. 2]{cao2008} uses a similar encoding structure as that of $\Rcal_{RTD}$ with two exceptions:
a) the binning is done sequentially rather than jointly as in $\Rcal_{RTD}$ leading to binning constraints (43)--(45) in \cite[Thm. 2]{cao2008}
 as opposed to \eqref{eq:our achievable region R0"}--\eqref{eq:our achievable region R1"+R2"} in Thm.\ref{thm:our achievable region}. Notable is that both schemes have adopted a Marton-like binning scheme at the cognitive transmitter, as first introduced in the context of the CIFC in \cite{cao2008}. b) While the cognitive messages are rate-split in identical fashions, the primary message is split into 2 parts in \cite[Thm. 2]{cao2008} ($R_1=R_{11}+R_{10}$, note the reversal of indices) while we explicitly split the primary message into three parts $R_2 = R_{2c}+R_{2pa}+R_{2pb}$.
We show that the region of \cite[Thm.2]{cao2008}, denoted as ${\cal R}_{CC} \subseteq {\cal R}_{RTD}$ in  two steps:

\noindent $\bullet$  We first show that we may WLOG set $U_{11} = \emptyset$ in \cite[Thm.2]{cao2008}, creating a new region $R_{CC}'$.

\noindent $\bullet$  We next make a correspondence between our RVs and those of \cite[Thm.2]{cao2008} and  obtain identical regions. 

We note that the primary and cognitive indices are permuted in \cite{cao2008}.

We first show that $U_{11}$ in \cite[Thm. 2]{cao2008} may be dropped WLOG. Consider the  region ${\cal R}_{CC}$ of \cite[Thm. 2]{cao2008}, defined as the union over all distributions $p_{U_{10},U_{11},V_{11},V_{20},V_{22},X_1,X_2}p_{Y_1,Y_2|X_1,X_2}$ of all rate tuples satisfying:
\eas{
R_1 &\leq I(Y_{1};V_{11},U_{11}, V_{20}, U_{10})
\label{eq:CC 37}\\
R_2 &\leq I(Y_2;  V_{20}, V_{22}|U_{10}) -I(V_{22}, V_{20}; U_{11}| U_{10})
\label{eq:CC 38}\\
R_1+R_2 &\leq I(Y_1; V_{11}, U_{11}| V_{20}, U_{10})+I(Y_2; V_{22},V_{20},U_{10})- I(V_{22};U_{11},V_{11}|V_{20},U_{10})
\label{eq:CC 39}\\
R_1+R_2 & \leq I(Y_1; V_{11}, U_{11},V_{20}, U_{10})+I(Y_2; V_{22}|V_{20},U_{10})- I(V_{22};U_{11},V_{11}|V_{20},U_{10})
\label{eq:CC 40}\\
2 R_2+R_1 &\leq I(Y_1;V_{11},U_{11},V_{20}|U_{10})+I(Y_2; V_{22}|V_{20},U_{10}) + I(Y_2;V_{20},V_{22},U_{10})\notag \\& \;\;\;\;  - I(V_{22};U_{11},V_{11}|V_{20},U_{10})-I(V_{22}, V_{20}; U_{11}| U_{10})
\label{eq:CC 41}
}{\label{eq:cao region FME}}

Now let ${\cal R}_{CC}'$ be the region obtained by setting
$U_{11}'=\emptyset$ and $V_{11}'=(V_{11},U_{11})$
while keeping all remaining RVs identical. Then ${\cal R}_{CC}'$ is the union over all distributions $p_{U_{10},V'_{11},V_{20},V_{22},X_1,X_2}p_{Y_1,Y_2|X_1,X_2}$, 
with $V_{11}'=(V_{11},U_{11})$ in $\Rcal_{CC}$,
 of all rate tuples satisfying:
\eas{
R_1 &\leq I(Y_{1};V_{11},U_{11}, V_{20}, U_{10})
\label{eq:CC 37p}\\
R_2 &\leq I(Y_2;  V_{20}, V_{22}|U_{10})
\label{eq:CC 38p}\\
R_1+R_2 &\leq I(Y_1; V_{11}, U_{11}| V_{20}, U_{10})+I(Y_2; V_{22},V_{20},U_{10})- I(V_{22};U_{11},V_{11}|V_{20},U_{10})
\label{eq:CC 39p}\\
R_1+R_2 &\leq I(Y_1; V_{11}, U_{11},V_{20}, U_{10})+I(Y_2; V_{22}|V_{20},U_{10})- I(V_{22};U_{11},V_{11}|V_{20},U_{10})
\label{eq:CC 40p}\\
2 R_2+R_1 &\leq I(Y_1;V_{11},U_{11},V_{20}|U_{10})+I(Y_2; V_{22}|V_{20},U_{10})+ I(Y_2;V_{20},V_{22},U_{10}) \notag \\& \;\;\;\; - I(V_{22};U_{11},V_{11}|V_{20},U_{10})
\label{eq:CC 41p}
}{\label{eq:cae FME 2}}
Comparing the two regions equation by equation, we see that
\begin{itemize}
  \item \eqref{eq:CC 37}= \eqref{eq:CC 37p}
  \item \eqref{eq:CC 38} $<$ \eqref{eq:CC 38p} as this choice of RVs sets the generally positive mutual information to 0
  \item \eqref{eq:CC 39}=\eqref{eq:CC 39p}
  \item \eqref{eq:CC 40}=\eqref{eq:CC 40p}
  \item \eqref{eq:CC 41} $<$ \eqref{eq:CC 41p} as this choice of RVs sets the generally positive mutual information to 0
\end{itemize}

\medskip

From the previous, we may set $U_{11} = \emptyset$ in  the region ${\cal R}_{CC}$ of \cite[Thm. 2]{cao2008} without loss of generality, obtaining the region ${\cal R}_{CC}'$ defined in \eqref{eq:CC 37p} -- \eqref{eq:CC 41p}.
We show that ${\cal R}_{CC}'$ may be obtained from the  region ${\cal R}_{RTD}$ with the assigment  of RVs, rates and binning rates
in Table \ref{tab:biao}. 


\begin{table}
\centering
\caption{Assignment of RVs of Section \ref{sec:biao} }\label{tab:biao}
\begin{tabular}{| lll |} \hline
 RV, rate of Theorem \ref{thm:our achievable region}  &  RV, rate of  \cite[Thm. 1]{DevroyeThesis} & Comments \\ \hline
$U_{2c}, R_{2c}$ &$ U_{10}, R_{10}$  & TX 2 $\goes$ RX 1, RX 2\\  
$X_{2}=U_{2c}$, $R_{2pa}=0$ & $U_{11}=\emptyset$, $R_{11}=0$      & TX 2 $\goes$ RX 2\\  
$U_{1c}, R_{1c}$ & $V_{20}, R_{20}$ & TX 1 $\goes$ RX 1, RX 2\\  
$U_{1pb}, R_{1pb}$ & $V_{22}, R_{22}$ & TX 1 $\goes$ RX 1\\  
$U_{2pb}, R_{2pb}$ &  $V_{11}$ &  TX 1 $\goes$  RX 2\\  
$ R_{1c}' $ &$L_{20}-R_{20}$ &\\
$ R_{1pb}' $&$L_{22}-R_{22}$ &\\
$ R_{2pb}' $&$L_{11}-R_{11}$ &\\
$X_1$                 & $X_2$               & \\
$X_2$                 & $X_1$               & \\
\hline
\end{tabular}
\end{table}


Evaluating  ${\cal R}_{CC}'$ defined by  \eqref{eq:CC 37p} -- \eqref{eq:CC 41p} with  the above assignment, translating all RVs into the notation used here, we obtain the region:
\pp{
R_{1c}' &\geq& 0 \\
R_{1pb}'+R_{2pb}'&\geq& I(U_{1pb}; U_{2pb} |U_{2c}, U_{1c}) \\
R_{2pb}+R_{2pb}' &\leq& I(Y_2; U_{2pb}|U_{2c}, U_{1c}) \\
R_{2pb}+R_{2pb}'+R_{1c}+R_{1c}' &\leq& I(Y_2; U_{1c}, U_{2pb}|U_{2c}) \\
R_{2pb}+R_{2pb}'+R_{1c}+R_{1c}'+R_{2c} &\leq& I(Y_2; U_{1c}, U_{2c}, U_{2pb}) \\
R_{1pb} +R_{1pb}' &\leq& I(Y_1;U_{1pb}| U_{2c}, U_{1c}) \\
R_{1pb}+R_{1pb}'+R_{1c}+R_{1c}' &\leq& I(Y_1;U_{1pb},U_{1c}| U_{2c}) \\
R_{1pb}+R_{1pb}'+R_{1c}+R_{1c}'+R_{2c} &\leq& I(Y_1;U_{1pb},U_{1c},U_{2c}) \\
}
Note that we may take binning rate equations $R_{1c}'\geq 0$ and $R_{1pb}'+R_{2pb}' \geq I(U_{1pb}; U_{2pb} |U_{2c}, U_{1c})$  to be equality without loss of generality - the largest region will take $R_{1c}', R_{1pb}', R_{2pb}'$ as small as possible.
The region ${\cal R}_{RTD}$ with $R_{2pa}=0$
\pp{
R_{1c}' &\geq& 0\\
R_{1c}'+R_{1pb}'&\geq& 0\\
R_{1c}'+R_{1pb}'+R_{2pb}' &\geq& I(U_{1pb}; U_{2pb} |U_{2c}, U_{1c}) \\
R_{2pb}+R_{2pb}' &\leq& I(Y_2;U_{2pb}|U_{2c}, U_{1c}) \\
R_{2pb}+R_{2pb}'+R_{1c}+R_{1c}' &\leq& I(Y_2; U_{1c}, U_{2pb}|U_{2c}) \\
R_{2pb}+R_{2pb}'+R_{1c}+R_{1c}'+R_{2c} &\leq& I(Y_2; U_{1c}, U_{2c}, U_{2pb}) \\
R_{1pb}+R_{1pb}' &\leq& I(Y_1;U_{1pb}| U_{2c}, U_{1c}) \\
R_{1pb}+R_{1pb}'+R_{1c}+R_{1c}' &\leq& I(Y_1;U_{1pb},U_{1c}| U_{2c}) \\
R_{1pb}+R_{1pb}'+R_{1c}+R_{1c}'+R_{2c} &\leq& I(Y_1;U_{1pb},U_{1c},U_{2c}) \\
}
For $R_{1c}'=0$ these two regions are identical, showing that ${\cal R}_{RTD}$ is surely no smaller than ${\cal R}_{CC}$.  For $R_{1c}'>0$, ${\cal R}_{RTD}$ , the binning rates of the region ${\cal R}_{RTD}$ are looser than the ones in ${\cal R}_{CC}$. This is probably due to the fact that the first one uses joint binning and latter one sequential binning.  Therefore ${\cal R}_{RTD}$ may produce rates larger than ${\cal R}_{CC}$. However, in general, no strict inclusion of ${\cal R}_{CC}$ in ${\cal R}_{RTD}$ has been shown.

\subsection{Containment of  \cite[Thm. 4.1]{jiang-achievable-BCCR} in ${\cal R}_{RTD}$ of Section \ref{sec:jiang BCCR}}
\label{sec:Jiang BCCR region}


In this scheme the common messages are created independently instead of having the common message from transmitter~1 being superposed to the common message from transmitter~2. The former choice introduces more rate constraints than the latter and allows us to show inclusion in $\Rcal_{RTD}$.

Again, following the argument of \cite[Appendix D]{willems1985discrete},
 we can show that  without loss of generality we can take $X_1$ and $X_2$ to be deterministic functions. With this consideration we can express the region of \cite[Thm. 4.1]{jiang-achievable-BCCR} as:
\begin{subequations}
\ea{
R_{22}' &\geq& I(W_2; V_1 ,X_1| U_1, U_2)
\label{Jiang 2}\\
R_{11}' + R_{22}' &\geq& I(W_2; W_1,V_1 ,X_1| U_1, U_2)
\label{Jiang 3}\\
R_{11}+R_{11}' &\leq& I(V_1,X_1, W_1; Y_1| U_1, U_2)
\label{Jiang 4}\\
R_{12}+R_{11}+R_{11}' &\leq& I(U_1, V_1, X_1 ,W_1 ; Y_1 | U_2)
\label{Jiang 5}\\
R_{21}+R_{11}+R_{11}' &\leq& I(U_2, V_1,X_1 ,W_1 ; Y_1 | U_1)
\label{Jiang 6}\\
R_{12}+R_{21}+R_{11}+R_{11}' &\leq& I(U_1, V_1, X_1 W_1, U_2; Y_1 )
\label{Jiang 7}\\
R_{22}+R_{22}' &\leq& I(W_2; Y_2 | U_1, U_2)
\label{Jiang 8}\\
R_{21}+R_{22}+R_{22}' &\leq& I(U_2,W_2; Y_2 | U_1)
\label{Jiang 9}\\
R_{12}+R_{22}+R_{22}' &\leq& I(U_1,W_2; Y_2 | U_2)
\label{Jiang 10}\\
R_{12}+R_{21}+R_{22}+R_{22}' &\leq& I(U_1,U_2,W_2; Y_2).
\label{Jiang 11}
}
\label{eq: Jiang-0}
\end{subequations}
taken over the union of all distributions
$$
p_{U_1}p_{V_1|U_1} p_{X_1|V_1,U_1}p_{U_2}p_{W_1,W_2| V_1,U_1, U_2} p_{X_0|W_1, W_2,V_1,U_1,U_2} p_{Y_1,Y_2|X_1,X_0}
$$
for
$(R_{11}',R_{22}',R_{11},R_{12},R_{21},R_{22})\in\RR^6_+.$

We can now eliminate one RV by  noticing that
\pp{
p_{U_1}p_{V_1|U_1} p_{X_1|V_1,U_1}p_{U_2}p_{W_1,W_2| V_1,U_1, U_2} p_{X_0|W_1, W_2,V_1,U_1,U_2} p_{Y_1,Y_2|X_1,X_0}\\
\lag =p_{U_1} p_{V_1,X_1|U_1}p_{U_2}p_{W_1,W_2| V_1,U_1, X_1,U_2} p_{X_0|W_1, W_2,V_1,U_1,X_1,U_2} p_{Y_1,Y_2|X_1,X_0},
}
and setting  $V_1'=[V_1,X_1]$, to obtain the region

\begin{subequations}
\ea{
R_{22}' &\geq& I(W_2; V_1'| U_1, U_2)
\label{Jiang 2}\\
R_{11}' + R_{22}' &\geq& I(W_2; W_1,V_1'| U_1, U_2)
\label{Jiang 3}\\
R_{11}+R_{11}' &\leq& I(V_1', W_1; Y_1| U_1, U_2)
\label{Jiang 4}\\
R_{12}+R_{11}+R_{11}' &\leq& I(U_1, V_1' ,W_1 ; Y_1 | U_2)
\label{Jiang 5}\\
R_{21}+R_{11}+R_{11}' &\leq& I(U_2, V_1' ,W_1 ; Y_1 | U_1)
\label{Jiang 6}\\
R_{12}+R_{21}+R_{11}+R_{11}' &\leq& I(U_1, V_1' W_1, U_2; Y_1 )
\label{Jiang 7}\\
R_{22}+R_{22}' &\leq& I(W_2; Y_2 | U_1, U_2)
\label{Jiang 8}\\
R_{21}+R_{22}+R_{22}' &\leq& I(U_2,W_2; Y_2 | U_1)
\label{Jiang 9}\\
R_{12}+R_{22}+R_{22}' &\leq& I(U_1,W_2; Y_2 | U_2)
\label{Jiang 10}\\
R_{12}+R_{21}+R_{22}+R_{22}' &\leq& I(U_1,U_2,W_2; Y_2)
\label{Jiang 11}
}
\label{eq: Jiang-0}
\end{subequations}
taken over the union of all distributions of the form
$$
p_{U_1}p_{V_1'|U_1} p_{U_2}p_{W_1,W_2| V_1',U_1, U_2} p_{X_0|W_1, W_2,V_1',U_1,U_2} p_{Y_1,Y_2|V_1',X_0}.
$$

We equate the RVs in the region of \cite{jiang-achievable-BCCR} with the RVs in Theorem \ref{thm:our achievable region} as in Table \ref{tab:Jiang BCCR region}.

\begin{table}
\centering
\caption{Assignment of RVs of Section \ref{sec:Jiang BCCR region} }\label{tab:Jiang BCCR region}
\begin{tabular}{| lll |} \hline
 RV, rate of Theorem \ref{thm:our achievable region}  &  RV, rate of  \cite[Thm. 1]{DevroyeThesis} & Comments \\ \hline
$U_{2c}, R_{2c}$ &$ U_{1}, R_{12}$  & TX 2 $\goes$ RX 1, RX 2\\  
$X_{2}, R_{2pa}$ & $V_{1}',R_{11}'$      & TX 2 $\goes$ RX 2\\  
$U_{1c}, R_{1c}$ & $U_{2}, R_{21}$ & TX 1 $\goes$ RX 1, RX 2\\  
$U_{1pb}, R_{1pb}$ & $W_{2}, R_{22}$ & TX 1 $\goes$ RX 1\\  
$U_{2pb}, R_{2pb}=0$ &  $W_{1}$ &  TX 1 $\goes$  RX 2\\  
$ R_{1c}' $ &$L_{20}-R_{20}$ &\\
$ R_{1pb}' $&$L_{11}-R_{11}$ &\\
$ R_{2pb}' $&$L_{22}-R_{22}$ &\\
$X_1$                 & $X_0$               & \\
$X_2$                 & $X_1$               & \\
\hline
\end{tabular}
\end{table}

With the substitutions of  Table \ref{tab:Jiang BCCR region} in the achievable rate region of  \reff{eq: Jiang-0}, we obtain the region
\begin{subequations}
\ea{
R_{1pb}' &\geq& I(U_{1pb}; X_2| U_{2c}, U_{1c})
\label{Jiang 1-0}\\
R_{1pb}' + R_{2pb}' &\geq &I(U_{1pb}; U_{2pb},X_2| U_{2c}, U_{1c})
\label{Jiang 1-1}\\
R_{2pa}+R_{2pb}' &\leq& I(X_2, U_{2pb}; Y_2| U_{2c}, U_{1c})
\label{Jiang 1-2}\\
R_{2c}+R_{2pa}+R_{2pb}' &\leq& I(U_{2c}, X_2 ,U_{2pb} ; Y_2 | U_{1c})
\label{Jiang 1-3}\\
R_{1c}+R_{2pa}+R_{2pb}' &\leq& I(U_{1c}, X_2 ,U_{2pb} ; Y_2 | U_{2c})
\label{Jiang 1-4}\\
R_{2c}+R_{1c}+R_{2pa}+R_{2pb}' &\leq& I(U_{2c}, X_2, U_{1c}, U_{1pb}; Y_2 )
\label{Jiang 1-5}\\
R_{1pb}+R_{1pb}' &\leq& I(U_{1pb}; Y_1 | U_{2c}, U_{1c})
\label{Jiang 1-6}\\
R_{1c}+R_{1pb}+R_{1pb}' &\leq& I(U_{1c},U_{1pb}; Y_1 | U_{2c})
\label{Jiang 1-7}\\
R_{2c}+R_{1pb}+R_{1pb}' &\leq& I(U_{2c},U_{1pb}; Y_1 | U_{1c})
\label{Jiang 1-8}\\
R_{2c}+R_{1c}+R_{1pb}+R_{1pb}' &\leq& I(U_{2c},U_{1c},U_{1pb}; Y_1).
\label{Jiang 1-9}
}
\label{eq: Jiang-1}
\end{subequations}
taken over the union of all distributions of the form
$$
p_{U_{1c}}p_{U_{2c}}p_{X_2|U_{2c}} p_{U_{1pb}, U_{2pb}| U_{1c}, U_{2c}, X_2} p_{X_1| U_{2c}, U_{1c},U_{1pb},U_{2pb}}.
$$
%
%
%
%
%
%
Set  $R_{2pb}=0$ and $R_{1c}'=I(U_{1c}; X_2| U_{2c})$ in the achievable scheme of Theorem \ref{thm:our achievable region} 
and consider the factorization of the remaining RVs
%
%
as in the scheme of \reff{eq: Jiang-1}, that is, according to
$$
p_{U_{1c}}p_{U_{2c}}p_{X_2|U_{2c}} p_{U_{1pb},U_{2pb}| U_{1c}, U_{2c}, X_2}p_{X_1| U_{2c}, X_2, U_{1c},U_{1pb},U_{2pb}}.
$$

With this factorization of the distributions, we obtain the achievable region

\begin{subequations}
\ea{
R_{1c}'&=&I(U_{1c}; X_2| U_{2c})\\
{R_{1pb}'}&\geq&{I(U_{1pb};X_2|U_{2c},U_{1c})}\label{us 1 - 0}\\
R_{1pb}'+R_{2pb}' &\geq& I(U_{1pb}; X_2, U_{2pb}| U_{2c},U_{1c})
\label{us 1-1}\\
R_{2pa} +R_{2pb}' &\leq& I(Y_2; X_2,U_{2pb}| U_{2c},U_{1c})        +I(U_{1c}; X_2| U_{2c})
\label{us 1-2}\\
R_{1c}+R_{2pa}+R_{2pb}' &\leq& I(Y_2; U_{1c},       X_2,U_{2pb}| U_{2c})
\label{us 1-3}\\
 R_{2c}+R_{1c}+R_{2pa}+R_{2pb}' &\leq& I(Y_2; U_{2pb},U_{1c},U_{2c},X_2)
 \label{us 1-4}\\
R_{1pb}     +R_{1pb}' &\leq& I(Y_1;               U_{1pb}|U_{2c},U_{1c})
\label{us 1-5}\\
R_{1c}+R_{1pb}+R_{1pb}' &\leq& I(Y_1;        U_{1c},U_{1pb}|U_{2c})
\label{us 1-6}\\
 R_{2c}+R_{1c}+R_{1pb}+R_{1pb}' &\leq& I(Y_1; U_{2c},U_{1c},U_{1pb})
 \label{us 1-7}
}
\label{eq: us}
\end{subequations}
Note that with this particular factorization we have that $I(U_{1c}; X_2|U_{2c})=0$,  since $X_2$ is conditionally independent of $U_{1c}$ given {$U_{2c}$.}

We now compare the region of \reff{eq: Jiang-1} and \reff{eq: us} for a fixed input distribution, equation by equation:
\pp{
\reff{us 1 - 0}=\reff{Jiang 1-0}\\
\reff{us 1-1}=\reff{Jiang 1-1}\\
\reff{us 1-2}=\reff{Jiang 1-2}\\
\reff{us 1-3}=\reff{Jiang 1-4}\\
\reff{us 1-4}=\reff{Jiang 1-5}\\
\reff{us 1-5}=\reff{Jiang 1-6}\\
\reff{us 1-6}={\reff{Jiang 1-7}}\\
\reff{us 1-7}=\reff{Jiang 1-9}\\
}
We see that {\reff{Jiang 1-3}} and { \reff{Jiang 1-8}} are extra bounds that further restrict the region in \cite {jiang-achievable-BCCR} to be contained in the region of Theorem \ref{thm:our achievable region}.




\end{document}